\pdfoutput=1
\documentclass[cernpreprint, atlasdraft=flase, texlive=2023, UKenglish, texmf, orcidlogo]{atlasdoc}

\usepackage{atlaspackage}
\usepackage{multirow}
\usepackage[backref=false]{atlasbiblatex}
 
\usepackage[BSM]{atlasphysics}

\graphicspath{{logos/}{figures/}}
 
\usepackage{ANA-EXOT-2019-05-PAPER-defs}

\AtlasTitle{Search for light long-lived neutral particles that decay to collimated pairs of leptons or light hadrons in $pp$ collisions at $\sqrt{s}=13$~TeV with the ATLAS detector}
 
\AtlasAbstract{
A search for light long-lived neutral particles with masses in the \textit{O}(MeV--GeV) range is presented. The analysis targets the production of long-lived dark photons in the decay of a Higgs boson produced via gluon--gluon fusion or in association with a $W$ boson. Events that contain displaced collimated Standard Model fermions reconstructed in the calorimeter or muon spectrometer are selected in 139~fb$^{-1}$ of $\sqrt{s} = 13$~TeV $pp$ collision data collected by the ATLAS detector at the LHC. Background estimates for contributions from Standard Model processes and instrumental effects are extracted from data. The observed event yields are consistent with the expected background. Exclusion limits are reported on the production cross-section times branching fraction as a function of the mean proper decay length $c\tau$ of the dark photon, or as a function of the dark-photon mass and kinetic mixing parameter that quantifies the coupling between the Standard Model and potential hidden (dark) sectors. A Higgs boson branching fraction above 1\% is excluded at 95\% CL for a Higgs boson decaying into two dark photons for dark-photon mean proper decay lengths between 10~mm and 250~mm and dark photons with masses between 0.4~GeV and 2~GeV.
}
\author{The ATLAS Collaboration}
 
\AtlasRefCode{EXOT-2019-05}

\PreprintIdNumber{CERN-EP-2022-054}

\AtlasJournal{JHEP}
\AtlasJournalRef{JHEP 06 (2023) 153}
\AtlasDOI{10.1007/JHEP06(2023)153}

\hypersetup{pdftitle={ATLAS document},pdfauthor={The ATLAS Collaboration}}
 
\begin{document}
 
\maketitle

\section{Introduction}
\label{sec:intro}
 
Dark sectors weakly coupled to the Standard Model (SM) are predicted in several theories of physics beyond the Standard Model (BSM)~\cite{ArkaniHamed:2008qp,Baumgart:2009tn,Falkowski:2014ffa,Curtin:2014cca}. Such theories often include a complex spectrum of new particles that form the dark sector and can couple to the SM only via specific portals. Depending on the structure of the dark sector and its coupling to the SM, unstable dark states may be produced at colliders, and could decay into SM particles with sizeable branching fractions.
 
This paper investigates the case where the SM and dark sectors communicate through the Higgs portal for the production of the BSM states and through a vector portal for the decay of the produced particles. In particular, a dark photon (\darky) is assumed to mix kinetically with the SM photon and decay into leptons and light quarks~\cite{Cheung:2009su,Falkowski:2010cm,Falkowski:2010gv}. The mass of the \darky and the kinetic mixing parameter ($\epsilon$), which is theoretically allowed to vary over a wide range of values, determines the lifetime of the dark photon.
This analysis focuses on small values of the kinetic mixing parameter, $\epsilon < 10^{-5}$, corresponding to \darky\ decays happening at a macroscopic distance from their production point, and a dark-photon mass in the \textit{O}(\MeV--\GeV) range. The decay branching fractions of a light dark photon that mixes kinetically with the SM photon depend on its mass \cite{Falkowski:2010cm, Meade:2009rb, Batell:2009yf}. Due to their small mass compared to the energy scale of the hard-scattering process, the dark photons are expected to be produced with large Lorentz boosts, resulting in collimated groups of fermions in a jet-like structure, referred to hereafter as dark-photon jets (\LJs).
 
\begin{figure}[ht!]
\centering
\subfloat[FRVZ 2\darky]{\includegraphics[width=.42\textwidth]{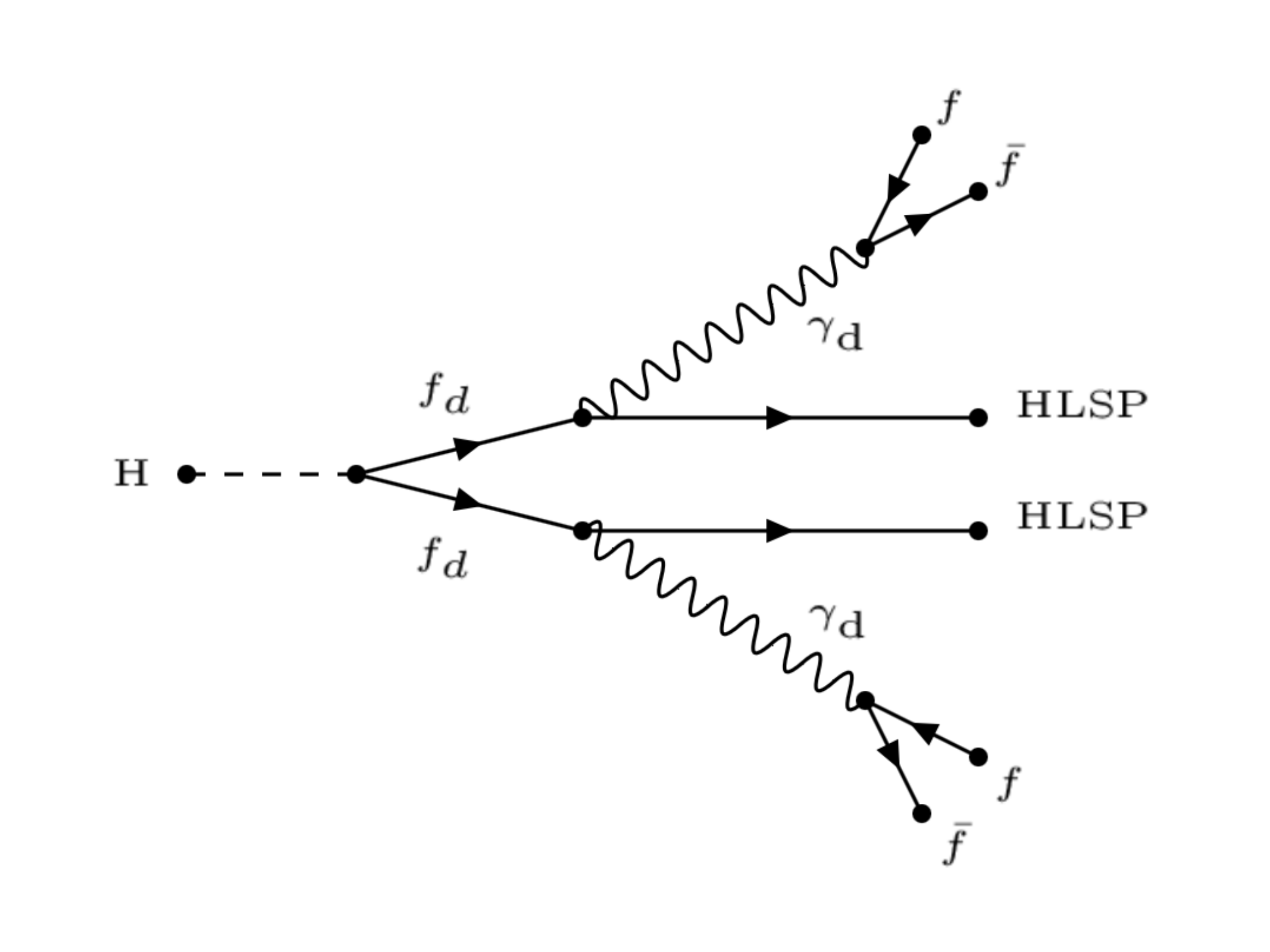}}
\subfloat[FRVZ 4\darky]{\includegraphics[width=.42\textwidth]{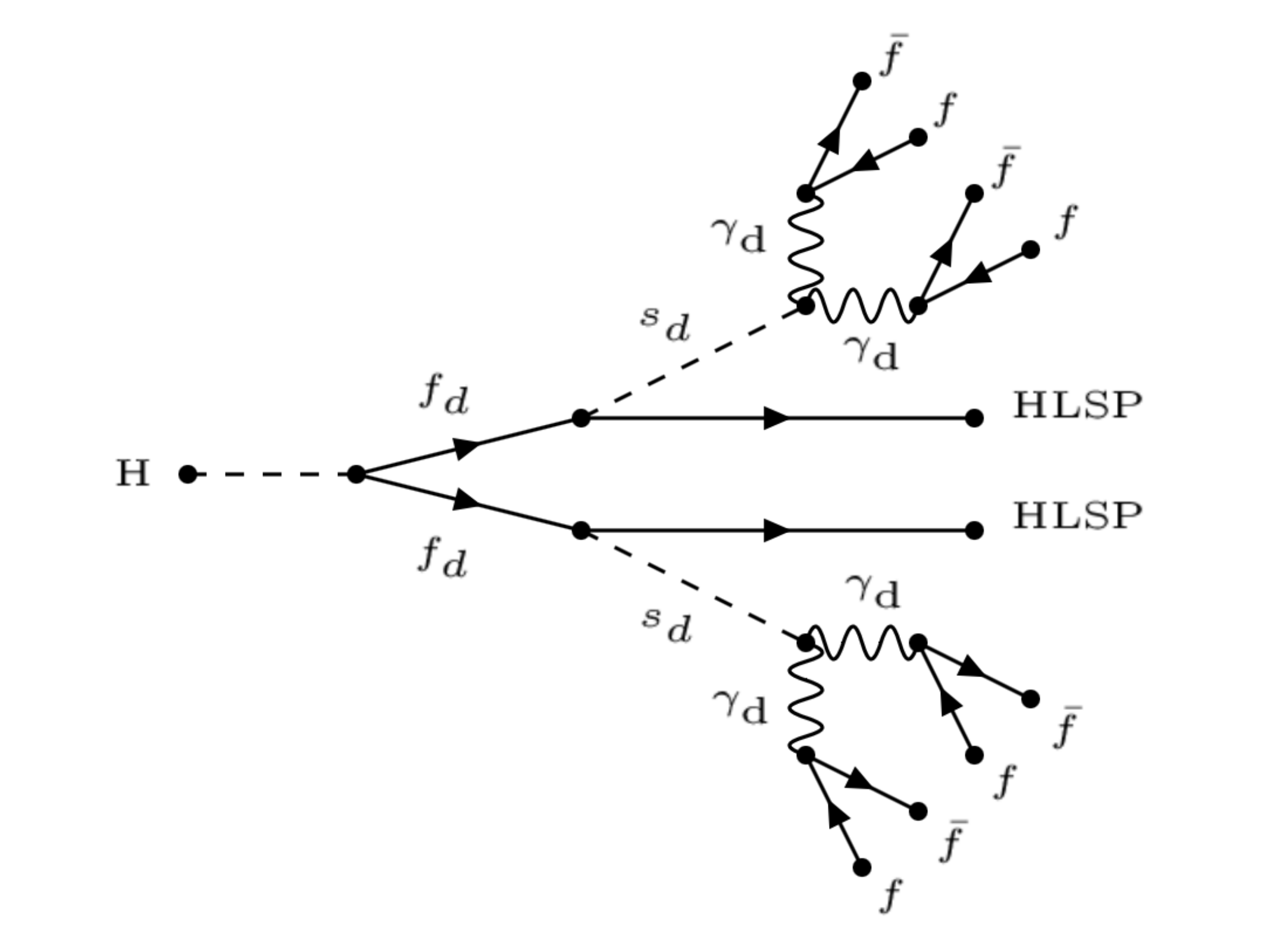}}\\
\subfloat[HAHM 2\darky]{\includegraphics[width=.42\textwidth]{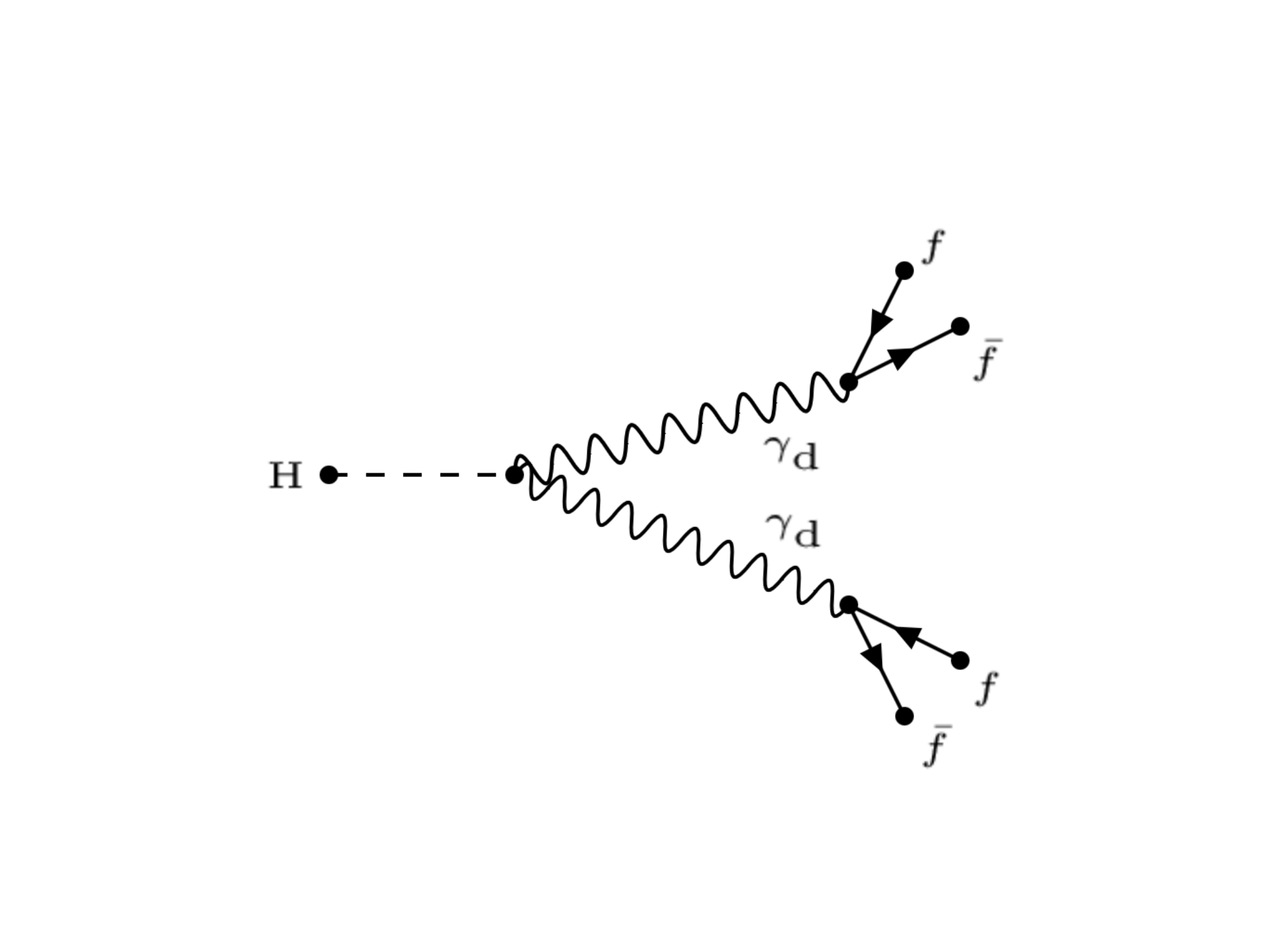}}
\caption{(a) FRVZ process with the dark fermion \fd decaying into a \darky and an HLSP.
(b) FRVZ process with the dark fermion \fd decaying into an HLSP and a dark scalar \sd that in turn decays into a pair of dark photons.
(c) HAHM process with the two \darky\ directly produced by the Higgs boson ($H$). The \darky\ decays into SM fermions, denoted by $f$ and $\bar{f}$. }
\label{fig:models}
\end{figure}

Two mutually exclusive search categories, referred to as \textit{gluon--gluon fusion selection} and \textit{WH associated production selection}, are designed to target the production of dark photons in these two Higgs boson ($H$) production modes. Two models are used for the optimisation of the event selections and the interpretation of the final results: the Falkowski--Ruderman--Volansky--Zupan (FRVZ) model \cite{Falkowski:2010cm, Falkowski:2010gv} and the Hidden Abelian Higgs Model (HAHM)~\cite{Curtin:2014cca}.
In the FRVZ model, a pair of dark fermions \fd is produced via a Higgs boson decay and can lead to final states with either two (Figure~\ref{fig:models}(a)) or four (Figure~\ref{fig:models}(b)) dark photons. In the two dark photon case, \fd decays into a dark photon and a stable dark fermion, which is assumed to be the hidden lightest stable particle (HLSP). Whereas in the four dark photon case, \fd decays into an HLSP and a dark scalar \sd that in turn decays into a pair of \darky.
The HAHM model predicts a direct decay of the Higgs boson into a pair of \darky, as shown in Figure~\ref{fig:models}(c).
The mean proper lifetime $\tau$ of the \darky\ is inversely related to the mass of the dark photon and to the square of the kinetic mixing parameter \cite{Cheung:2009qd, Batell:2009yf}.

In general, dark-sector radiation~\cite{Strassler:2006im} can produce extra dark photons proportionally to the size of the dark gauge coupling $\alpha_\textrm{d}$~\cite{Cheung:2009su}. In this work a dark coupling $\alpha_\textrm{d} \lesssim 0.01$ is assumed, leading to no significant additional \darky emission.
This signal feature is not explicitly exploited and sensitivity is retained for new physics scenarios with multiple \darky emission, although no specific interpretations are given for these scenarios.

The search for displaced \LJs presented in this paper uses the dataset collected at the Large Hadron Collider (LHC) by the ATLAS detector during 2015--2018 in proton--proton ($pp$) collisions at a centre-of-mass energy $\sqrt{s}=13$~\TeV, corresponding to an integrated luminosity of 139~\ifb. Previous searches for displaced \LJs were performed by ATLAS using proton--proton collision data at lower centre-of-mass energies~\cite{EXOT-2011-12, EXOT-2013-22} or a partial dataset corresponding to 36.1~\ifb\ of proton--proton collision data at a centre-of-mass energy of $\sqrt{s} = 13$~\TeV~\cite{EXOT-2017-28}. The results are complementary to those from related ATLAS searches for prompt \LJs~\cite{EXOT-2011-01,EXOT-2012-16,EXOT-2014-09}, which probed higher values of $\epsilon$, and for displaced dimuon vertices using 13~\TeV\ data~\cite{EXOT-2017-03}, which probed higher dark-photon mass values. Related searches for dark photons were conducted by the CDF and D0 collaborations at the Tevatron~\cite{Aaltonen:2012pu,Abazov:2009hn,Abazov:2010uc} and by the CMS~\cite{CMS-EXO-11-013,CMS-EXO-12-012,CMS-HIG-13-010,CMS-EXO-19-007} and LHCb~\cite{Aaij:2015tna,LHCb:2019vmc} collaborations at the LHC.
Additional constraints on scenarios with dark photons are extracted from, e.g., beam-dump and fixed-target experiments~\cite{Blumlein:2011mv,Bjorken:2009mm,Bross:1989mp,Merkel:2011ze,Adlarson:2013eza,Abrahamyan:2011gv,Reece:2009un,Blumlein:2013cua,Gninenko:2012eq,Essig:2010gu,Agakishiev:2013fwl}, $e^{+}e^{-}$ colliders~\cite{Archilli:2011zc,Babusci:2012cr,Aubert:2009cp,Lees:2014xha,Lees:2015rxq,TheBelle:2015mwa,Won:2016pjz,Ablikim:2018bhf,Ablikim:2017aab}, electron and muon anomalous magnetic moment measurements~\cite{Pospelov:2008zw,Davoudiasl:2012ig,Endo:2012hp} and astrophysical observations~\cite{Chang:2018rso,Dreiner:2013mua}.
Given the various constraints, a displaced dark-photon decay with a kinetic mixing parameter $\epsilon < 10^{-5}$ is allowed for \darky masses greater than 10~\MeV.
In this paper, the sensitivity to displaced \LJs is significantly higher than in the previous ATLAS search~\cite{EXOT-2017-28}, due to the higher integrated luminosity of the dataset and to new analysis methods.
The new analysis methods include updated signal-region selection criteria and \LJ\ reconstruction techniques.
Multivariate techniques based on convolutional neural networks, exploiting the three-dimensional representations of calorimeter energy deposits associated with jets, and a dense neural network, using muon track information, are employed to indentify \LJ\ candidates.
The addition of a new set of event selection criteria, targeting events where the dark photons arise from the decay of a scalar particle produced in association with a $W$ boson, significantly increases the sensitivity in scenarios with \darky\ mean proper decay lengths $c\tau$ below (above) 3 (40)~mm.


\FloatBarrier

\section{ATLAS detector}
\label{sec:detector}
\newcommand{\AtlasCoordFootnote}{
ATLAS uses a right-handed coordinate system with its origin at the nominal interaction point (IP) in the centre of the detector and the \(z\)-axis along the beam pipe. The \(x\)-axis points from the IP to the centre of the LHC ring, and the \(y\)-axis points upwards. Cylindrical coordinates \((r,\phi)\) are used in the transverse plane, \(\phi\) being the azimuthal angle around the \(z\)-axis. The pseudorapidity is defined in terms of the polar angle \(\theta\) as \(\eta = -\ln \tan(\theta/2)\), and the rapidity is defined as \(y = (1/2) [(E+p_z)/(E-p_z)]\).}
 
The ATLAS detector~\cite{PERF-2007-01} at the LHC covers nearly the entire solid angle around the collision point.\footnote{\AtlasCoordFootnote}
It consists of an inner tracking detector (ID) surrounded by a thin superconducting solenoid, electromagnetic (EM) and hadron (HCAL) calorimeters, and a muon spectrometer (MS) incorporating three large superconducting air-core toroidal magnets.
 
The inner tracking detector covers the pseudorapidity range \(|\eta| < 2.5\).
It consists of silicon pixel, silicon microstrip, and transition radiation tracking detectors.
 
The calorimeter system covers the pseudorapidity range \(|\eta| < 4.9\).
Within the region \(|\eta|< 3.2\), electromagnetic calorimetry is provided by barrel and endcap high-granularity lead/liquid-argon (LAr) calorimeters, with an additional thin LAr presampler covering \(|\eta| < 1.8\) to correct for energy loss in material upstream of the calorimeters.
Hadron calorimetry is provided by the steel/scintillator-tile calorimeter, segmented into a central barrel and two extended-barrel cylindrical structures within \(|\eta| < 1.7\), and two copper/LAr hadron endcap calorimeters.
The solid angle coverage is completed with forward copper/LAr and tungsten/LAr calorimeter modules optimised for electromagnetic and hadronic energy measurements respectively.
Overall, the calorimeter system is highly segmented in shower depth, with a total of seven sampling layers in the barrel region and eight in the endcap regions.
 
The muon spectrometer comprises separate trigger and high-precision tracking chambers measuring the deflection of muons in a magnetic field generated by the superconducting air-core toroidal magnets.
The field integral of the toroids ranges between \num{2.0} and \SI{6.0}{\tesla\metre} across most of the detector.
A set of precision chambers covers the region \(|\eta| < 2.7\) with three layers of monitored drift tubes, complemented by cathode-strip chambers in the forward region, where the background is highest.
The muon trigger system covers the range \(|\eta| < 2.4\) with resistive-plate chambers in the barrel, and thin-gap chambers in the endcap regions.
 
A two-level trigger system is used to select events~\cite{TRIG-2016-01}. The first-level (L1) trigger is implemented in hardware and uses a subset of the detector information to accept events at a rate of at most \SI{100}{\kHz}. This is followed by a software-based high-level trigger (HLT) that reduces the accepted event rate to \SI{1}{\kHz} on average, depending on the data-taking conditions.
 
An extensive software suite~\cite{ATL-SOFT-PUB-2021-001} is used in the reconstruction and analysis of real and simulated data, in detector operations, and in the trigger and data acquisition systems of the experiment.


\FloatBarrier

\section{Data and simulated event samples}
\label{sec:samples}
 
The data were collected by the ATLAS detector during Run~2 of the LHC (2015--2018) with a peak instantaneous luminosity of $2.1\times 10^{34}$~cm$^{-2}$s$^{-1}$, resulting in a mean number of $pp$ interactions per bunch crossing of $\langle \mu \rangle = 34$.
Data quality requirements~\cite{DAPR-2018-01} are applied to ensure that all subdetectors were operating normally, and that the LHC beams were in stable-collision mode. The integrated luminosity of the resulting data sample is 139~\ifb. The data were collected using a set of triggers specifically designed to target the displaced \darky decays, described in detail in Section~\ref{sec:ggF}, or single-lepton triggers, with requirements on the identification, isolation, and \pT\ of the leptons to maintain efficiency across the full momentum range while controlling the trigger rates~\cite{TRIG-2018-01,TRIG-2018-05}.
 
During $pp$ collisions data-taking the LHC circulates two counter-rotating proton beams constructed from bunches of protons. However, following LHC injection not all bunch slots are filled with protons, with the number of unfilled bunches depending on the LHC filling scheme~\cite{Evans:2008zzb}. An empty bunch crossing takes place when neither beam is filled with protons and each empty bunch is separated from filled bunches by at least five empty bunches on each side. A dataset enriched in cosmic-ray muon background is collected during empty bunch crossings (the `cosmic dataset') and used for the background estimation.
 
Monte Carlo (MC) simulated event samples are used to model the BSM signals. Signal samples modelling the production of dark photons via a 125~\GeV\ Higgs portal were generated with \MGNLO[2.2.3]~\cite{Alwall:2014hca} interfaced to \Pythia[8.186]~\cite{Sjostrand:2007gs} for the parton showering (PS) and hadronisation. The matrix-element calculation was performed at tree level. The parton distribution function (PDF) set used for the generation is~\NNPDF[2.3lo]~\cite{Ball:2012cx}. Higgs boson production via gluon--gluon fusion and in association with a $W$ boson is included. The predicted Standard Model cross-sections for these two processes, assuming $m_{H}=$~125.09~\GeV, are respectively 48.61~pb~\cite{Cepeda:2019klc,Anastasiou:2016cez} and 1.369~pb~\cite{deFlorian:2016spz,Brein:2003wg}. A second set of signal samples was generated, modelling the production via gluon--gluon fusion of a high-mass (800~\GeV) Higgs-like scalar mediator with the same decay modes as in the 125~\GeV\ mass case.
Effects of higher-order QCD corrections on the \pT\ of the Higgs boson, evaluated using a reweighting procedure~\cite{ATL-PHYS-PUB-2021-020}, change the signal selection efficiency by less than the MC statistical accuracy and are therefore neglected.
A dark photon with a mass $m_{\gamma_\textrm{d}}$ up to a few \GeV\ that mixes kinetically with the SM photon will decay into leptons or light quarks, with branching fractions that depend on its mass \cite{Falkowski:2010cm, Meade:2009rb, Batell:2009yf}. Taking as an example a dark-photon mass of 0.4~\GeV, the \darky~decay branching ratios are expected to be 45$\%~\ee$, 45$\%~\mu^+\mu^-$, and 10$\%~q\bar{q}$~\cite{Falkowski:2010cm}. The mean proper decay length $c\tau$ of the \darky is treated as a free parameter. In the generated samples, \ctau was chosen such that, accounting for the boost of the \darky, a large fraction of the decays occur inside the sensitive ATLAS detector volume (i.e.\ up to 7~m in radius and 13~m along the $z$-axis from the centre of the detector).
The decays of the Higgs boson into dark photons through dark fermions or directly into two dark photons were simulated at matrix-element level during the generation.
In the FRVZ model, the mass of \fd was chosen to be small relative to the Higgs boson mass, and far from the kinematic threshold at $m_{\mathrm{HLSP}} + m_{\gamma_{\mathrm{d}}} = m_{f_{\mathrm{d}}}$. The values of these masses have a negligible impact on the analysis results. In the HAHM, the Higgs boson decays directly into two dark photons.
 
The SM background is estimated using data-driven techniques, with MC simulated events used to aid in the validation, the evaluation of uncertainties, and the training of dedicated multivariate classifiers. SM  processes that could be potential sources of background include multi-jet, \textit{W}+jets, \textit{Z}+jets, \ttbar, single-top-quark, \textit{WW}, \textit{WZ}, and \textit{ZZ} production.
The multi-jet samples were generated with \PYTHIA[8.210]~\cite{Sjostrand:2014zea} using the same set of tuned parameter values (tune) and PDF as for the signal samples.
Samples of \textit{W}+jets, \textit{Z}+jets, \textit{WW}, \textit{WZ}, and \textit{ZZ} events were generated using \Sherpa[2.2.1]~\cite{Gleisberg:2008ta} with the \NNPDF[3.0nnlo]~\cite{Ball:2014uwa} PDF set. Single-top and \ttbar MC samples were generated using \POWHEGBOX[v2]~\cite{Frixione:2007vw} and \PYTHIA[8.230] with the A14 tune~\cite{ATL-PHYS-PUB-2014-021} for parton showering and hadronisation, and the \NNPDF[2.3lo] set of PDFs.
 
Finally, MC samples of $J/\psi \to \mu\mu$ events were generated and used to evaluate systematic uncertainties for muon trigger and reconstruction efficiencies. The MC samples were generated using \textsc{Pythia8+Photos++}~\cite{Golonka:2005pn} with the A14 tune for parton showering and hadronisation, and the CTEQ6L1~\cite{Gao:2013xoa,Pumplin:2002vw} PDF set.
 
All generated Monte Carlo events were processed through a full simulation of the ATLAS detector geometry and response~\cite{SOFT-2010-01} using the \GEANT~\cite{Agostinelli:2002hh} toolkit. The simulation includes multiple $pp$ interactions per bunch crossing (pile-up), as well as the detector response to interactions in bunch crossings before and after the one producing the hard interaction. In order to model the effects of pile-up, simulated inclusive $pp$ events were overlaid on each generated hard-scatter event and reweighted to match the conditions of the 2015--2018 data sample. The inclusive $pp$ events were simulated with \PYTHIA[8.210] with the A3 tune~\cite{ATL-PHYS-PUB-2016-017} and the~\NNPDF[2.3lo] set of PDFs.


\FloatBarrier

\section{Event reconstruction}
\label{sec:dpjreco}
Candidate events are required to have a reconstructed vertex~\cite{ATL-PHYS-PUB-2015-026} with at least two associated tracks with transverse momentum (\pT) larger than 500~\MeV\ that are consistent with originating from the beam collision region in the $x$--$y$ plane. The primary vertex in the event is the vertex with the highest sum of squared transverse momenta of associated tracks.
 
Electron candidates are reconstructed from isolated electromagnetic calorimeter energy deposits matched to ID tracks. They are required to have $|\eta|<2.47$, a transverse momentum $\pT>\SI{20}{\GeV}$, and to satisfy the `TightLHElectron' requirement defined in Ref.~\cite{ATL-PHYS-PUB-2015-027}, which is based on a likelihood evaluated using measurements of shower shapes in the calorimeter and track properties in the ID as input variables. Candidates within the transition region between the barrel and endcap electromagnetic calorimeters, $1.37 < |\eta| < 1.52$, are not considered.
 
Muon candidates are reconstructed in the region $|\eta|<2.5$ from MS tracks matching ID tracks.
They are required to have $\pT>\SI{20}{\GeV}$ and satisfy the `medium' identification requirements defined in Ref.~\cite{ATLAS-CONF-2018-023}. These requirements are based on the number of hits in the different ID and MS subsystems, and on the ratio  of the charge and momentum ($q/p$) measured in the ID and MS divided by the sum in quadrature of their corresponding uncertainties.
 
The tracks associated with the electron and muon candidates are required to have a significance of the transverse impact parameter,\footnote{The transverse impact parameter, $d_0$, is defined as the distance of closest approach of a track to the beam-line, measured in the transverse plane. The longitudinal impact parameter, $z_0$, corresponds to the $z$-coordinate distance between the point along the track at which the transverse impact parameter is defined and the primary vertex.} $d_0$, of $\vert d_0\vert/\sigma(d_0) < 5$ for electrons and $\vert d_0\vert/\sigma(d_0) < 3$ for muons, and a longitudinal impact parameter, $z_0$, satisfying $\vert z_0 \sin\theta\vert < 0.5$~mm.
 
Isolation criteria are applied to electrons and muons. The scalar sum of the \pt of ID tracks within a variable-size cone around the electron (muon), must be less than 15\% of the lepton \pt, excluding tracks associated with the lepton. The track isolation cone size for electrons (muons) \DRf\ is given by the smaller of $\Delta R = \SI{10}{~GeV}/\pt$ and $\Delta R = 0.2\,(0.3)$. In addition, the sum of the transverse energy of clusters of calorimeter cells~\cite{PERF-2014-07} in a cone of $\Delta R = 0.2$ around the electron (muon) must be less than 20\% (30\%) of the lepton \pt, excluding clusters associated with the lepton.
 
Jets are reconstructed from three-dimensional energy clusters in the calorimeter~\cite{ATL-PHYS-PUB-2015-036} using the anti-$k_t$ jet clustering algorithm~\cite{Cacciari:2008gp,Fastjet} with a radius parameter $R=0.4$. Only jet candidates with $\pT>\SI{20}{\GeV}$ and $|\eta|<4.9$ are considered.
Jets are calibrated using MC simulation, with corrections obtained from in situ techniques~\cite{PERF-2016-04}. To reduce the effects of pile-up, jets with $\pt<\SI{120}{\GeV}$ and $|\eta|<2.5$ are required to have a significant fraction of their associated tracks compatible with originating from the primary vertex, as defined by the jet vertex tagger~\cite{PERF-2014-03} (JVT). This requirement reduces the fraction of jets from pile-up to 1\%, with an efficiency for hard-scatter jets of about 90\%.
Jets not satisfying basic quality criteria designed to reject detector noise and non-collision backgrounds~\cite{ATLAS-CONF-2015-029} are then discarded.
 
Jets containing $b$-hadrons ($b$-tagging) are identified with 
a multivariate discriminant that makes use of track properties~\cite{ATL-PHYS-PUB-2017-013, FTAG-2018-01}.
Jets are considered to be $b$-tagged if they fulfil a requirement that has 70\% average efficiency for jets containing $b$-hadrons in simulated $\ttbar$ events.
The rejection factors for light-quark and gluon jets, jets containing $c$-hadrons, and hadronically decaying $\tau$-leptons in simulated $\ttbar$ events are approximately 301, 38, and 8, respectively.
 
Simulated events are corrected for differences from collision data in $b$-tagging efficiencies and $b$-tagging mis-tag rates~\cite{FTAG-2018-01, ATLAS-CONF-2018-001, ATLAS-CONF-2018-006}. Corrections are also applied to account for minor differences between data and MC simulation in the single-lepton trigger, reconstruction, identification and isolation efficiencies~\cite{EGAM-2018-01,MUON-2018-03}.
 
Jet candidates within an angular distance $\Delta R' =\sqrt{(\Delta y)^2+(\Delta\phi)^2} = 0.2$ of a lepton candidate are discarded.  Remaining lepton candidates within $\Delta R'= \operatorname{min}\{0.4, 0.04 + \pt(\mu)/(\SI{10}{\GeV})\}$ of a jet are then discarded to suppress bottom and charm hadron decays. When considering muons, if the jet has fewer than three associated tracks, the muon is retained and the jet is discarded instead to avoid inefficiencies for high-energy muons undergoing significant energy loss in the calorimeter. Finally, any electron candidate sharing an ID track with a remaining muon candidate is also removed.
 
The missing transverse momentum vector \ptmiss, whose magnitude is denoted by \MET, is defined as the negative vector sum of the transverse momenta of all identified electrons, muons and jets, plus an additional soft term. The soft term is constructed from all tracks that originate from the primary vertex but are not associated with any identified lepton or jet. In this way, the \MET is adjusted for the best calibration of leptons and jets, while contributions from pile-up interactions are suppressed through their exclusion from the soft term~\cite{ATL-PHYS-PUB-2015-027,ATLAS-CONF-2018-023}.

A displaced \darky\ decay producing a pair of muons is expected to leave two or more collimated stand-alone MS tracks, referred to as a muonic dark-photon jet \mLJ.
Stand-alone MS tracks~\cite{PERF-2015-10} are reconstructed in the pseudorapidity region $| \eta | \geq 0.1$ and formed by requiring at least two matched segments in the MS, and are fit with a primary vertex constraint.
Candidates with pseudorapidity in the range $1.0 \leq | \eta | \leq 1.1$ are rejected to avoid the transition region of the MS between barrel and endcap. Moreover, only stand-alone MS tracks in the pseudorapidity interval $|\eta|<2.4$, corresponding to the ID coverage, are selected to allow an isolation variable based on ID tracks to be computed. Stand-alone MS tracks are required to not match any prompt muon candidate, in order to discard muons coming from the main interaction vertex.
 
A \darky\ decaying into a displaced electron or quark pair leads to energy deposits in the calorimeters reconstructed as a single jet with low EM fraction (EMF), defined as the ratio of the energy deposited in the EM calorimeter to the total jet energy. Jets with EMF below 0.4 are referred to as calorimeter dark-photon jet \hLJ. MC simulations show that \LJs containing two dark photons both decaying into an electron or quark pair are reconstructed as a single jet. Low-EM-fraction jets are reconstructed and calibrated with the same algorithms as described previously. They are, however, only considered if they have $\pT>\SI{20}{\GeV}$ and lie within $|\eta|<2.5$. Furthermore, to retain high efficiency for the targeted signals, they are required to satisfy quality criteria looser than those in the main jet selection and no JVT selection is applied. To avoid selecting events where most of the energy associated with a jet could have been produced by localised noise, events are rejected if the leading jet has ${>}90$\% of its energy associated with a single constituent cluster or layer within the LAr calorimeter. Potential background from noise bursts~\cite{LARG-2013-01} in the LAr calorimeter is rejected via a veto on events where the leading jet's largest energy deposit is located in the EM calorimeter endcap, where most noise bursts occur.
Noise-induced jets in the hadron calorimeters are removed by imposing the \textit{BadLoose} cleaning selection, described in Ref.~\cite{ATLAS-CONF-2015-029}, without the cuts on the fraction of jet energy deposited within the electromagnetic calorimeter and the jet charged-fraction.
These cleaning requirements reject approximately 0.8\% of low-EM-fraction jets in the signal samples used in this analysis. Finally, the time $t_{\mathrm{\hLJ}}$ associated with a \hLJ, measured as the energy-weighted average of the timing for each calorimeter cell related to the jet and corrected by the corresponding time-of-flight from the interaction point, is required to fall within a window of 4~ns around zero. CaloDPJs from cosmic-ray muons and beam-induced backgrounds (BIB) would have a different jet-timing distribution than collision products.
 
Dedicated algorithms for the identification of \LJs have been developed to target \darky\ decays into \mLJs\ or \hLJs. These algorithms are described in Sections~\ref{sec:muonconst} and~\ref{sec:hadconst} respectively.
 
\subsection{Muonic dark-photon jets}
\label{sec:muonconst}
 
Muonic dark-photon jets are reconstructed using the Cambridge--Aachen clustering algorithm~\cite{Dokshitzer:1997in} that combines all the selected stand-alone MS tracks lying within a cone of fixed size in ($\eta$, $\phi$) space. The algorithm starts from the highest-\pT\ stand-alone MS track and searches for additional stand-alone MS tracks within a \delr~=~0.4 cone around the initial track's momentum vector. If a second stand-alone MS track is found in the cone, the axis of the cone is rotated to the vector sum of the momenta of the two tracks, and the search is repeated until no additional tracks are found in the cone. The \mLJs are required to have at least two MS tracks and are discarded if a jet is found within \delr~=~0.4 of a \mLJ, to ensure orthogonality between reconstructed \LJ types.
 
Cosmic-ray muons that cross the detector in time coincidence with a $pp$ interaction constitute the main source of background to the \mLJ. The cosmic dataset is used to study this background and to train a dense neural network (DNN), referred to as the cosmic-ray tagger, to discriminate signal \mLJs\ from the \mLJ candidates that originate from the cosmic-ray background. For optimal training, a balanced mixture of all available MC signal samples is used. The DNN is implemented using Keras with the Tensorflow backend~\cite{tensorflow2015-whitepaper} and it is trained to classify each stand-alone MS track constituting a \mLJ using the following quantities: the longitudinal impact parameter $z_{0}$, the track angular direction (in $\eta$ and $\phi$), and the timing measurements from the MS. The time measurements in different stations of the MS, when available, allow the muon's direction of flight to be identified. A parametric training method is used to only consider the timing measurements when they are available. The neural network has three dense hidden layers, with 32, 64 and 128 neurons respectively. An output layer with a sigmoidal activation function returns a binary classification score between 0 and 1, which can be interpreted as the probability for a \mLJ constituent track to originate from a \darky\ decay, and example distributions are shown in Figure~\ref{fig:muDPJtaggerscore}. The figure also illustrates the small dependency on the specific choice of signal model. A \mLJ is accepted if all its stand-alone MS track constituents have a cosmic-ray tagger output score >~0.5. This selection was optimised to retain a high signal efficiency: signal \mLJs\ are selected with an efficiency above 95\% for transverse decay lengths \lxy up to 5~m and for \darky\ \pT\ larger than 20~\GeV\ and with a background rejection of 90\%.
 
\begin{figure}[h]
\centering
\includegraphics[width=0.49\textwidth]{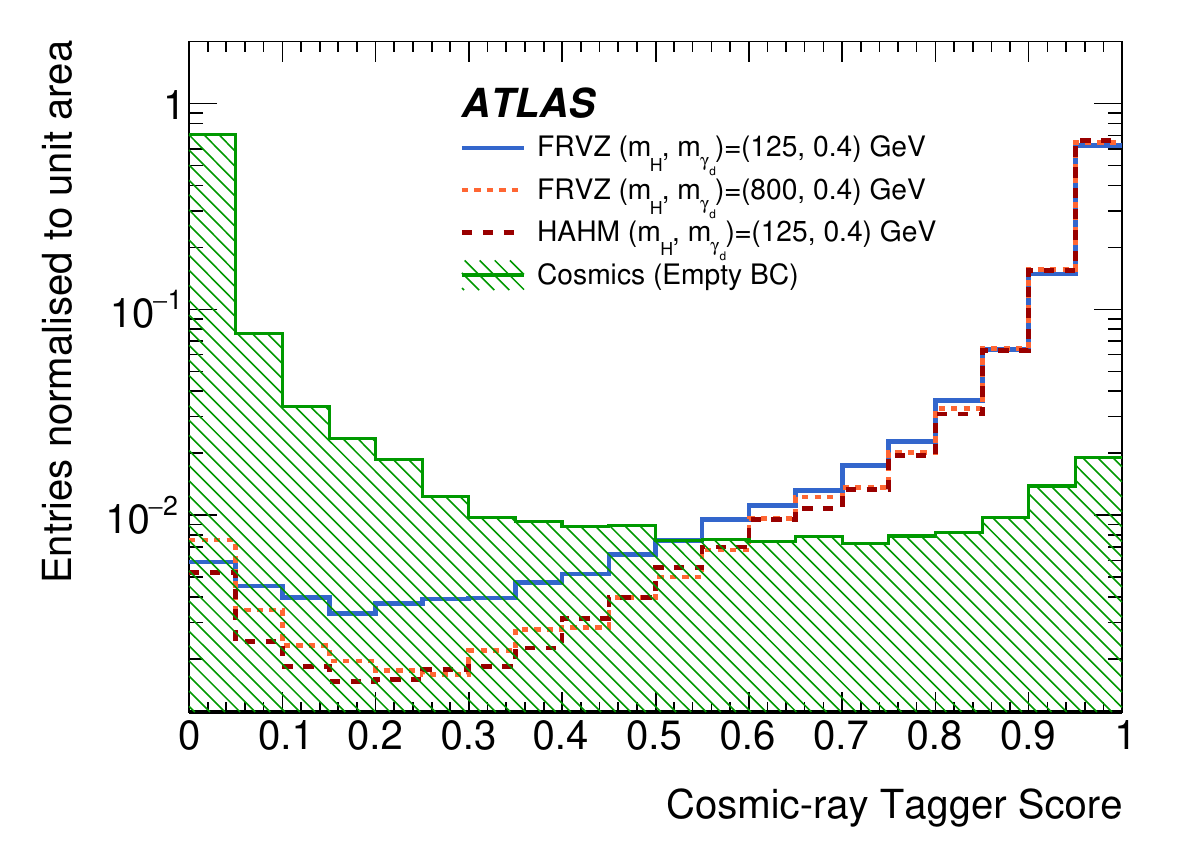}
\caption{The distribution of the cosmic-ray tagger output score is shown for a few signal scenarios (lines) and for data from the cosmic dataset (shaded area).}
\label{fig:muDPJtaggerscore}
\end{figure}
 
The \mLJ reconstruction efficiency is shown in Figure~\ref{fig:mu_reco_Lxy_Lz}, for \mLJ objects that satisfy the cosmic-ray tagger selection, as a function of the \lxy and transverse momentum of the dark photon in a few benchmark signal scenarios.
The efficiency mainly depends on the opening angle between the muons from the \darky\ decay, thus models with a small generated \ctau and a large dark-photon boost will be more efficiently reconstructed.
A drop in efficiency is expected for \darky decays which occur after the middle layer of the MS (6~m radius in the barrel region), where muons can no longer be reconstructed.
 
\begin{figure}[h]
\centering
\subfloat[]{\includegraphics[width=0.49\textwidth]{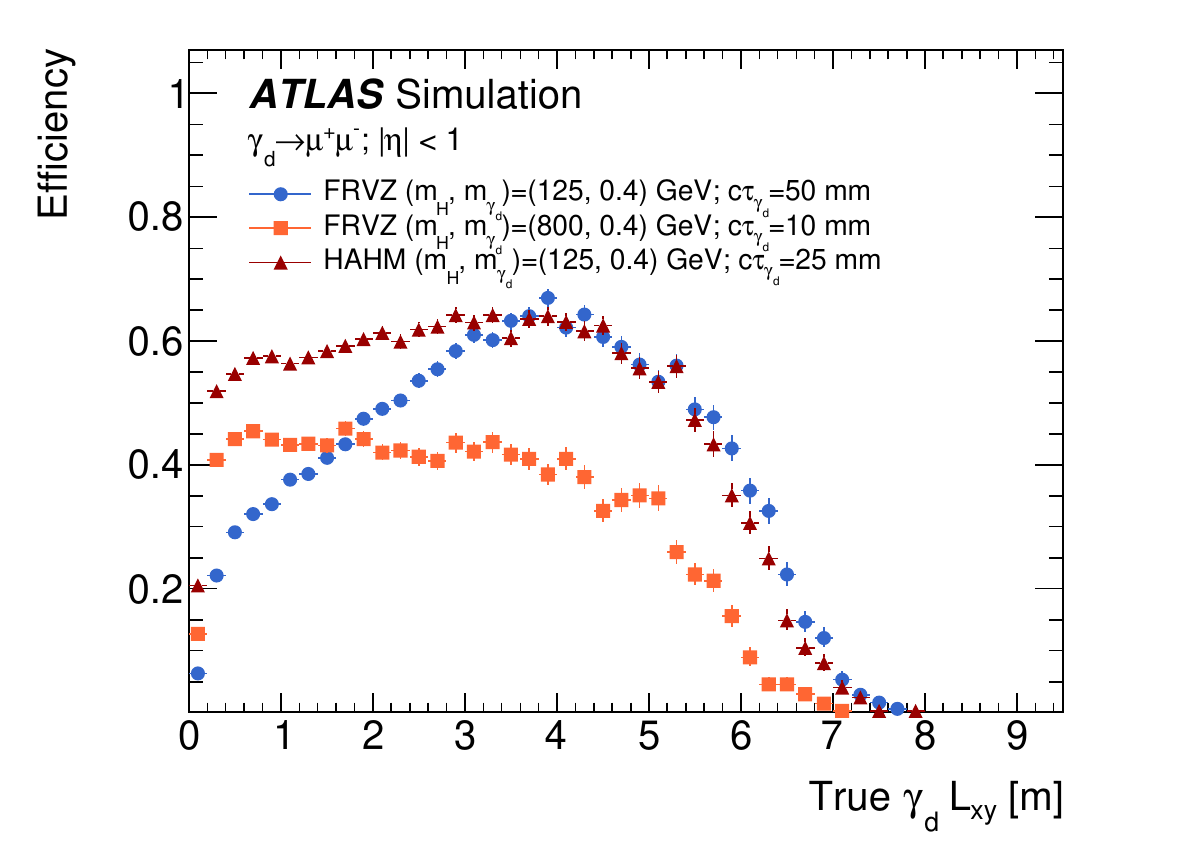}}
\subfloat[]{\includegraphics[width=0.49\textwidth]{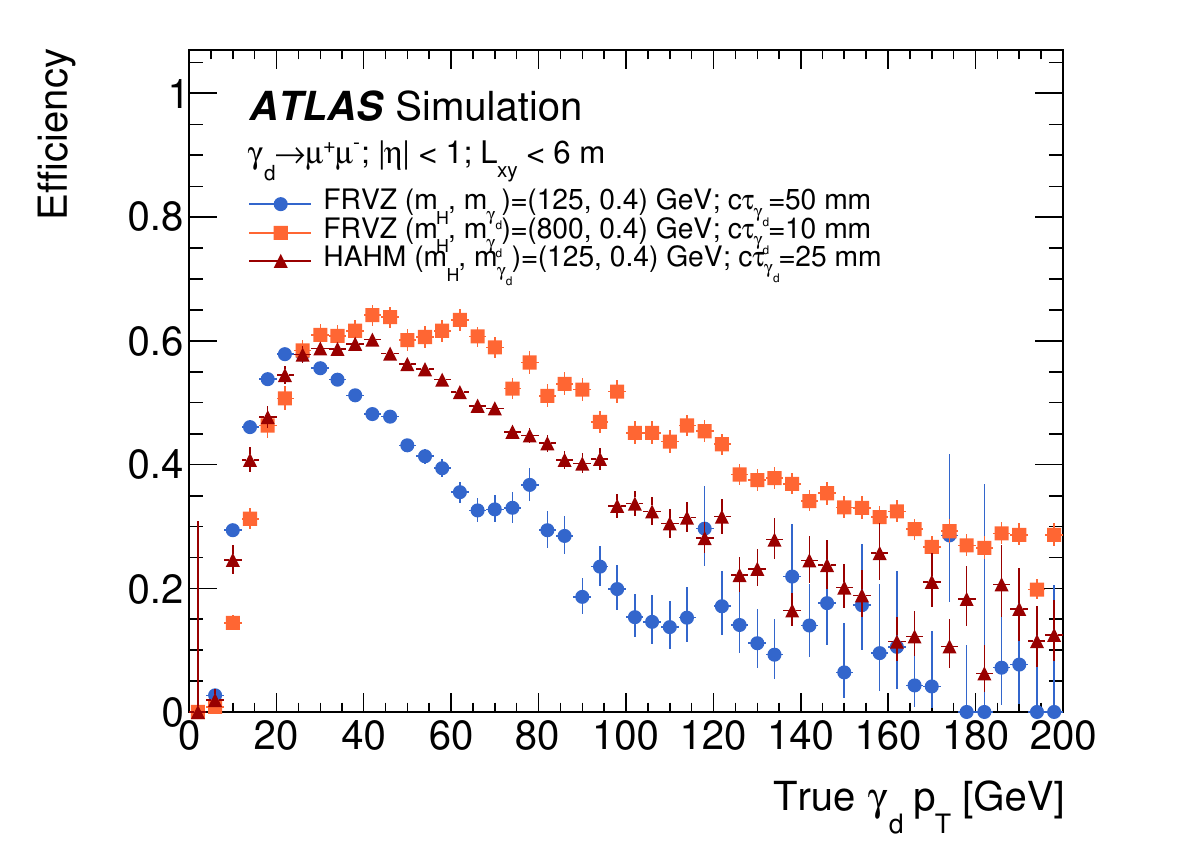}}
\caption{The reconstruction efficiency for \mLJ objects satisfying the cosmic-ray tagger selection produced in the decay of a \darky\ into a muon pair. Figure (a) shows the reconstruction efficiency for \darky with $0<|\eta|<1$ as a function of the transverse decay length \lxy, input to the MC generation and referred to as true \darky. Figure (b) shows the reconstruction efficiency for \darky\ with $0<|\eta|<1$ as a function of the  \darky\ transverse momentum in events where the \darky \lxy is below 6~m.}
\label{fig:mu_reco_Lxy_Lz}
\end{figure}
 
\subsection{Calorimeter dark-photon jets}
\label{sec:hadconst}
 
All the low-EM-fraction jets satisfying the selection described in Section~\ref{sec:dpjreco} are considered as calorimeter dark-photon-jet candidates. Candidates in the transition region between the barrel calorimeters and the endcap cryostat are removed by requiring the fraction of energy in the Tile Gap scintillators to be less than 10\% of the total jet energy.
 
The main source of background for \hLJs is jet production in multi-jet events. In order to reduce this background, a dedicated discriminator (\hLJtagger), based on a convolutional neural network implemented using Keras with the Tensorflow backend, is used to assign a score to each \hLJ in the event. The training of the neural network is based on simulated events and exploits \hLJs reconstructed in signal and multi-jet MC events. The \hLJtagger inputs are three-dimensional representations of energy deposits associated with the jet. The energy deposits are defined by collections of calorimeter cell clusters used in jet reconstruction~\cite{PERF-2014-07}. Each collection has $(\eta,\phi)$ coordinates and holds information about the total amount of energy deposited in each of the calorimeter samplings.
 
The $(\eta,\phi)$ space around the jet axis is divided into a $15\times 15$ grid, centred on the jet axis and corresponding to an $\eta\times\phi = 0.9\times0.9$ area, so that each cell cluster within this 2D grid corresponds to a $0.06\times0.06$ portion of the $(\eta,\phi)$ space. A third axis is added to this grid, to take into account the EM and HCAL calorimeters' samplings as an additional coordinate. The resulting 3D grids are composed of cell clusters which contain the total  energy released at the corresponding $(\eta,\phi)$ coordinates and calorimeter sampling. In order to exploit the full calorimeter volume, three 3D grids are produced: one accounting for the barrel samplings, one for the tile calorimeter's extended barrel and one for the endcap. These are then used as input to a convolutional neural network. The output layer of the neural network has a sigmoidal activation function resulting in a binary classification in the [0, 1] range, as shown in Figure~\ref{fig:hadtaggerscores}(a).
 
CaloDPJs are defined to be candidates if they have a QCD tagger score larger than 0.5. This selection was optimised to retain a high signal efficiency and corresponds to a background rejection of approximately 94\%. The \hLJtagger has a signal efficiency above 95\% for \lxy above 2.5~m, which decreases to about 60\% for \darky decaying within the ID. The efficiency was also found to be independent of the \darky\ \pT\ for a \pT\ larger than 20~\GeV.
 
Muons arising from BIB and crossing the detector longitudinally, at radial distances $r > 2\ \si{\meter}$, can deposit energy in the HCAL by radiative losses, which can be reconstructed as \hLJs owing to the resulting low EM fraction. To reduce the residual contamination from misidentified \hLJs from BIB, a dedicated per-jet tagger (\hLJBIBtagger) was developed. This tagger uses the same strategy as the \hLJtagger, exploiting the topology of the energy deposits to classify the candidate jets.
 
\begin{figure}[h]
\centering
\subfloat[]{\includegraphics[width=0.49\textwidth]{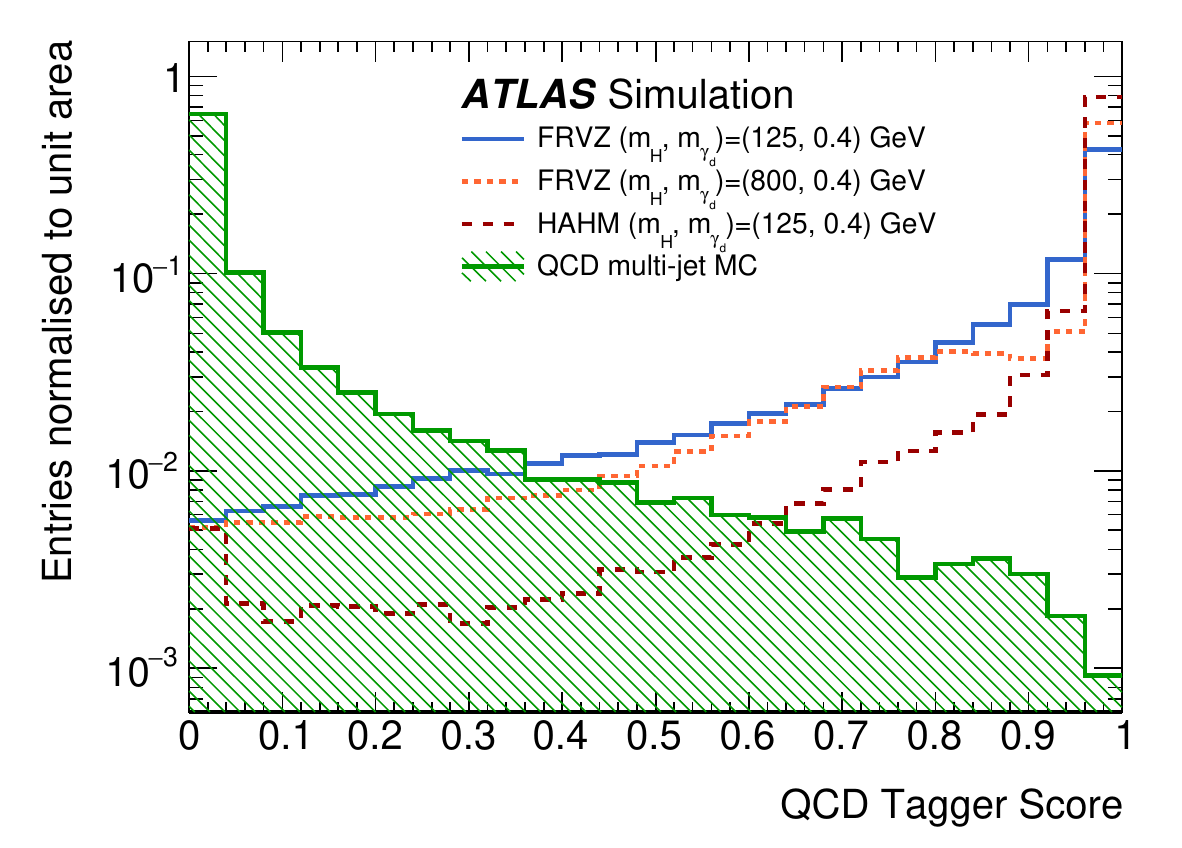}}
\subfloat[]{\includegraphics[width=0.49\textwidth]{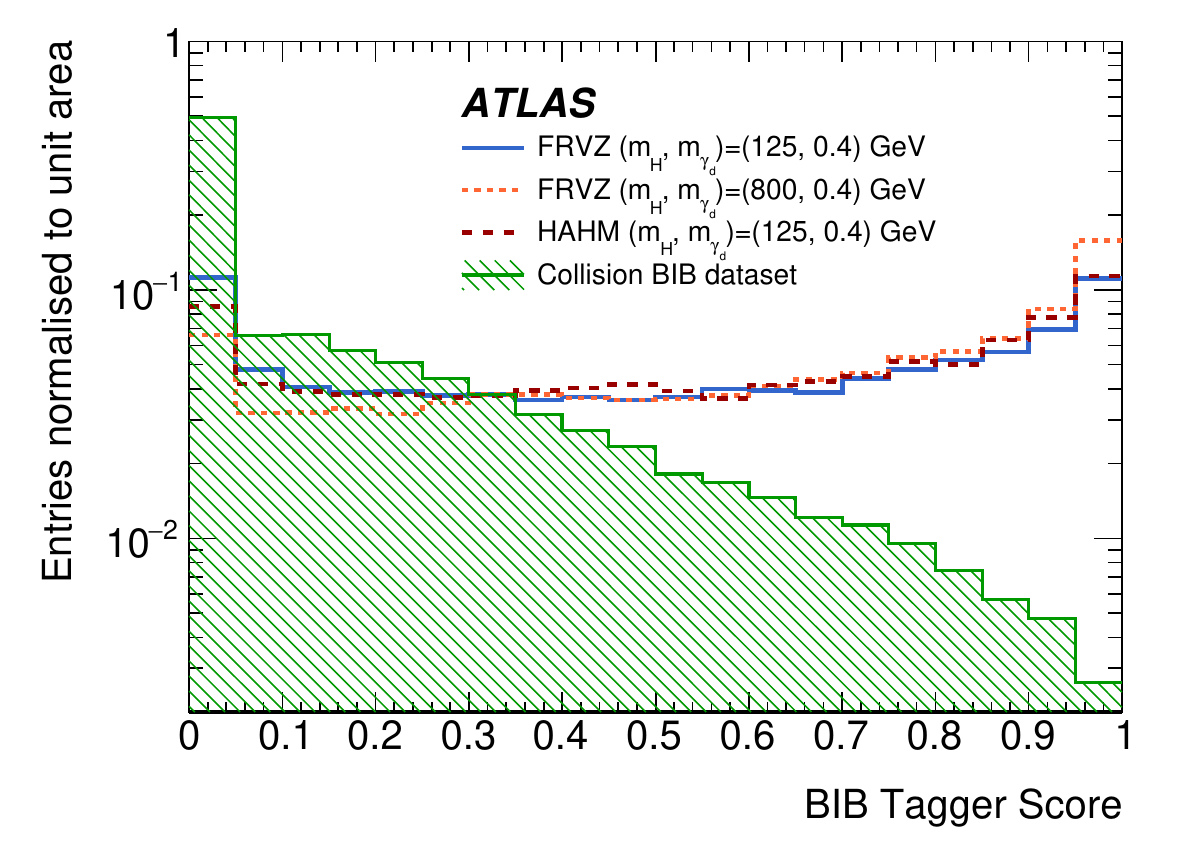}}
\caption{The output score for the taggers used on calorimeter dark-photon jets. Figure (a) shows the QCD tagger score for a few signal scenarios and simulated multi-jet events. Figure (b) shows the BIB tagger score for the same signal scenarios and BIB-enriched data.}
\label{fig:hadtaggerscores}
\end{figure}
 
The \hLJBIBtagger shares the network architecture of the \hLJtagger. To remove any possible energy dependence and rely only on geometrical information, the input tensors are preprocessed by substituting the cluster energies with a fixed value. The jet $(\eta,\phi)$ coordinates are passed as inputs to the dense neural network, in addition to the convolution layers' output. CaloDPJs from a BIB-enriched data selection and from all available MC signal samples are used as input for the training. The BIB-enriched data selection uses a dedicated calorimeter trigger exploiting cell timing and position to select BIB events. This trigger's performance was validated on BIB events occurring during empty bunch crossings. CaloDPJs are accepted if the output score of the \hLJBIBtagger, shown in Figure~\ref{fig:hadtaggerscores}(b), is larger than 0.2, corresponding to a BIB rejection rate of 68\%. This value is chosen in order to keep the signal efficiency above 80\%, irrespective of the signal scenario, for events entering the analysis selections.
 
The \hLJ reconstruction efficiency is shown in Figure \ref{fig:had_reco_Lxy_Lz} for \hLJ candidates that satisfy the QCD tagger selection but no BIB tagger requirement, as a function of the transverse decay length and transverse momentum, for several benchmark signal samples. A \darky decaying into hadrons or electrons is reconstructed only within the HCAL volume (at radii of 2.28--4.25~m in the barrel region).
 
\begin{figure}[h]
\centering
\subfloat[]{\includegraphics[width=0.49\textwidth]{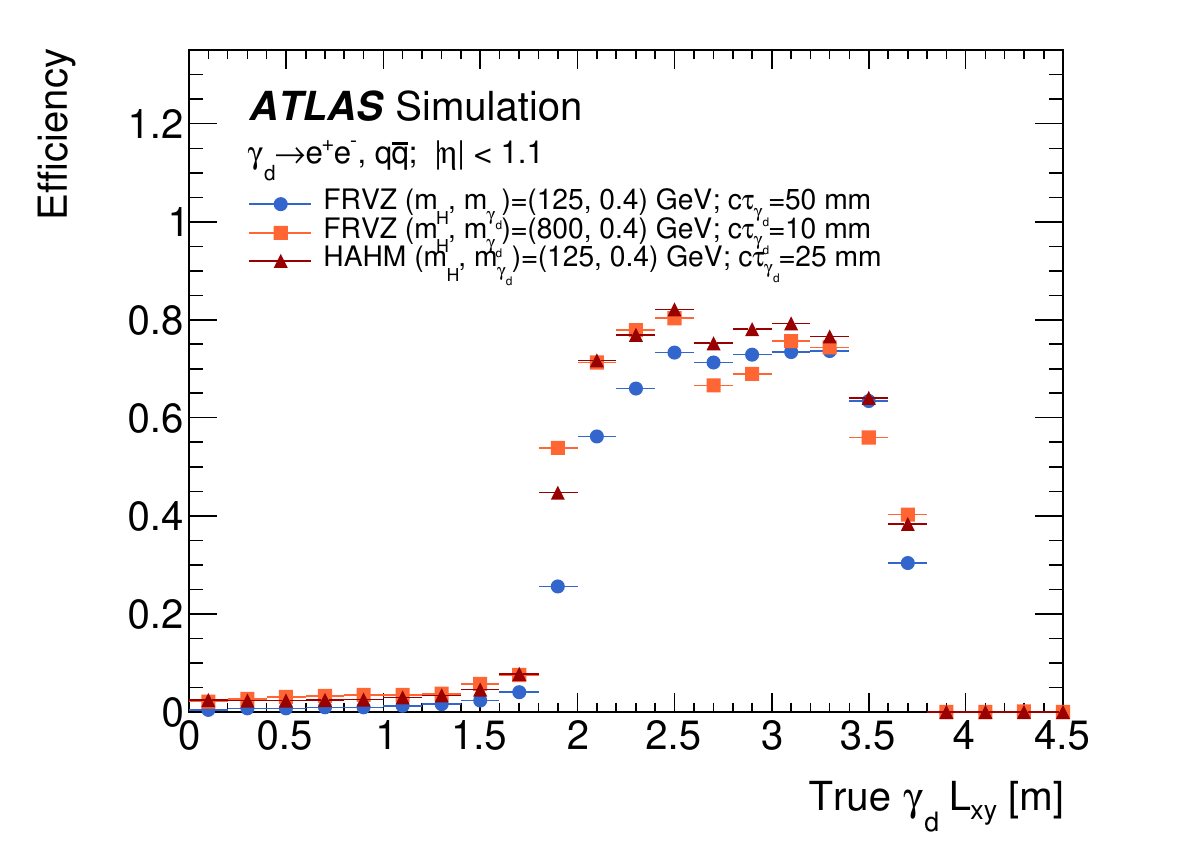}}
\subfloat[]{\includegraphics[width=0.49\textwidth]{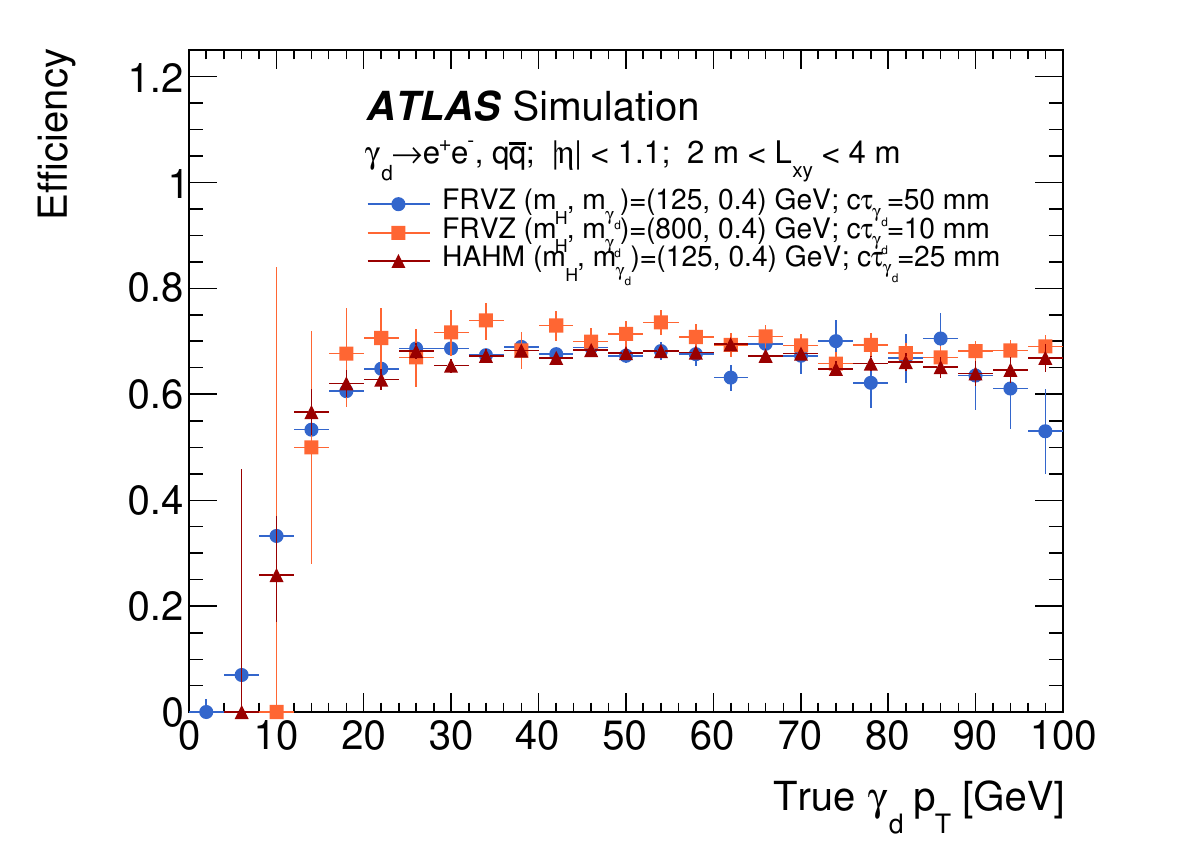}}
\caption{The reconstruction efficiency for \hLJs produced by the decay of \darky into \epem or \qqbar. Figure (a) shows the reconstruction efficiency for \darky with $0<|\eta|<1.1$ as a function of the transverse decay length \lxy, input to the MC generation and referred to as true \darky. The efficiency drop at 2.5~m corresponds to the end of the first layer of the HCAL. Figure (b) shows the reconstruction efficiency for \darky with $0<|\eta|<1.1$ as a function of the transverse momentum in events where the \darky \lxy is between 2~m and 4~m.}
\label{fig:had_reco_Lxy_Lz}
\end{figure}


\FloatBarrier

\section{Event selection and background estimation}
\label{sec:analysis}
 
The events are classified into the two exclusive search categories, gluon--gluon fusion (ggF) and \textit{WH} associated production, based on their charged-lepton multiplicity. The ggF category targets the Higgs boson production mode where events have no charged-lepton candidates satisfying the selections described in Section~\ref{sec:dpjreco}. The \textit{WH} category targets Higgs boson production in association with a $W$ boson and requires exactly one charged lepton.
At least one \LJ satisfying the selection criteria described in Section~\ref{sec:dpjreco} is required in the selected events. If more than two \LJs are reconstructed, the one with the highest transverse momentum, called the `leading \LJ', and the one farthest in \delphi\ from the leading one, called the `far \LJ', are used to classify the event.
Each search category further separates the events into different orthogonal search channels based on the numbers of \mLJs and \hLJs, resulting in a total of six signal regions (SRs) that were optimised for the best discovery sensitivity and are described in detail in Section~\ref{sec:ggF} and Section~\ref{sec:Wh}.
 
The main sources of background for the \LJ signals are cosmic-ray muons for the \mLJs and jet production for both the \hLJs and \mLJs. Cosmic-ray muon background for the \mLJs is strongly reduced after the event selection, and the main residual background is found to consist of hadronic or electromagnetic showers which reach the MS after not being fully contained in the calorimeter (punch-through jets) and originate from multi-jet events.
A subdominant background for both \LJ types is BIB. All background contributions are estimated from data. For the different decay modes considered, dedicated sets of the following discriminating variables are used to separate the BSM signal from the backgrounds or in the definition of regions used to aid the estimation of the background contributions:
\begin{itemize}
\item $m_{\textrm{jj}}$: the invariant mass of the two leading jets.
\item $|\Delta t_{\textrm{\hLJs}}|$: the absolute time difference between a pair of \hLJs. This quantity is useful for further rejecting contributions from cosmic-ray muons and BIB, as these \hLJ candidates do not originate from a single interaction vertex.
\item JVT: the JVT score of a \hLJ can be used to reject candidates that are likely to originate from the primary proton--proton interaction vertex.
\item $\Delta \phi_{\textrm{DPJ}}$: the azimuthal angle between the leading \LJ and far \LJ. Signal events are expected to contain anti-aligned \LJs.
\item \dpjiso: the scalar sum of the transverse momenta of all tracks within a $\Delta R=0.5$ cone around the direction of the \LJ\ momentum vector. Displaced \LJs are expected to have very little nearby track activity in the ID.
\item \maxiso: the largest of the \dpjiso\ values for the two \LJs in an event.
\item \prodcalo: the product of the QCD tagger scores of the two \LJs, or the single QCD tagger score when only one \hLJ is available.
\item $m_{\textrm T}$: the transverse mass in $WH$ events, computed as $m_{\textrm T} = \sqrt{2 p_\textrm{T}^{\ell}\met(1 - \cos \Delta\phi)}$ where $\Delta\phi$ is the azimuthal angle between the missing transverse momentum vector and the lepton, is exploited to reduce possible background contributions from multi-jet processes with a misidentified or non-prompt lepton.
\item \hLJ width ($w$): the \pT-weighted sum of the $\Delta R$ between each energy cluster of the jet and the jet axis. Jets from \LJs\ are expected to be narrower, on average, than ordinary jets since they are produced just before or inside the calorimeters.
\item \mindphi: smallest azimuthal angular separation between a selected \LJ\ and the \ptmiss vector.
\item \minhtag: the minimum QCD tagger score, computed for up to two \hLJs in an event.
\end{itemize}
 
The detailed selection requirements, together with the dedicated methods for the estimation of the residual background contributions, are described in Section~\ref{sec:ggF} for the ggF selection and in Section~\ref{sec:Wh} for the \textit{WH} selection.


\FloatBarrier

\subsection{Gluon--gluon fusion selection}
\label{sec:ggF}
 
In this selection, events are accepted if they satisfy one or more of three dedicated triggers targeting displaced objects. The trigger strategy is analogous to that described in Ref.~\cite{EXOT-2017-28}. It comprises two MS-based triggers and one calorimeter-based trigger.
 
The first MS-based trigger, referred to as tri-muon MS-only~\cite{TRIG-2018-01}, accepts events with at least three muons with $\pT \geq 6$~\GeV\ reconstructed using only MS information. The second MS-based trigger, referred to as muon narrow-scan~\cite{TRIG-2012-02}, requires a muon candidate from the L1 trigger with $\pT \ge 20$~\GeV\ to be confirmed by the HLT using only MS information. At the HLT a `scan' is then performed in a cone of $\Delta R = 0.5$ around this muon, looking for a second muon reconstructed using only MS information. The \pT requirement on the second muon was increased from 6~\GeV\ to 15~\GeV\ during the course of the 2015--2016 data taking and kept constant afterwards. Both muons were required to be not matched to any track in the ID, and track isolation was required for the leading muon. During the 2017--2018 data taking, the narrow-scan trigger was extended to take advantage of the L1 topological trigger~\cite{TRIG-2019-02} by requiring partially matched topological items. The leading L1 muon candidate was then combined with an HLT object: either an unmatched muon, a jet, or transverse momentum imbalance. The last two were added to target events with one \mLJ and one \hLJ.
The narrow-scan triggers have efficiencies of 75\% and 40\%, independent of the dark photon mass, for the 2015--2016 and 2017--2018 versions, respectively, when considering events with \darky decays into muon pairs with $|\eta|<1.0$ and \lxy below 6~m.
 
A L1 calorimeter trigger called the CalRatio trigger~\cite{TRIG-2012-02} is designed to select narrow jets produced in the hadronic calorimeter.
During 2015--2016, the CalRatio L1 trigger selected narrow jets with transverse energy $E_{\mathrm T}>60$~\GeV\ in a $0.2\times0.2$~$(\Delta\eta\times\Delta\phi)$ region of the EM and HCAL.
During 2017--2018 data taking, the CalRatio trigger instead used a L1 topological trigger~\cite{TRIG-2019-02} to select jets in the HCAL that have $E_{\mathrm T}>30$~\GeV\ and are isolated from energy deposits in the EM calorimeter with $\ET>3$~\GeV\ within $\Delta R = 0.2$ around the most energetic HCAL energy deposit.
At the HLT, jets from both of these L1 selections were required to have $E_{\mathrm T} \geq 30$~\GeV, $|\eta|<2.4$ and EM fraction~$< 0.06$. These jets were also required to have no tracks with $p_\mathrm{T} \geq 2$~\GeV\ within $\Delta R = 0.2$ of the jet axis, and to satisfy BIB-suppression requirements on calorimeter cell timing and position.
The CalRatio triggers used in 2015--2016 and 2017--2018 are found to have constant efficiencies of approximately 5\% and 20\%, respectively, for FRVZ signal events where the \darky is produced in the decay of a 125~\GeV\ Higgs boson and decays into \epem or \qqbar within $|\eta|<1.0$. These efficiencies rise to 25\% and 30\%, on average, for signal events in HAHM models, or to 55\% and 45\%, on average, when assuming \darky\ production via an 800~\GeV\ scalar mediator.
 
Events where the two leading jets have an invariant mass $m_{\textrm{jj}}>1$~\TeV\ and $\MET>225$~\GeV\ are removed to allow for a statistically independent study of vector-boson fusion (VBF) production modes in the future.
 
The events satisfying the requirements listed above are further separated into orthogonal search channels based on the \LJ types. Events with pairs of \mLJs (\hLJs) are targeted by \SRggFmm\ (\SRggFhh), while events with one \LJ\ of each kind are targeted by \SRggFhm. Each of these SRs includes additional selection requirements, as summarised in Table~\ref{tab:SRdefggF} together with the triggers employed.
Events with a single reconstructed \LJ are not considered for the ggF selection as the constraints they provide are not competitive due to the large background yields.
The impact of beam-induced background is suppressed by requiring \hLJ candidates to pass the BIB tagger selection and, only in the \SRggFhh where two \hLJs are reconstructed, to have a time difference $|\Delta t_{\textrm{\hLJs}}|$ smaller than $2.5$~ns.
In \SRggFhh\ and \SRggFhm, the leading and far \LJs\ are further required to have an azimuthal angular difference $\Delta \phi_{\textrm{DPJ}}$ larger than 0.5.
Multi-jet background events are rejected by requiring the \LJs to satisfy ID isolation criteria. \LJ candidates are required to have a \dpjiso below 4.5~\GeV.
 
\begin{table}[!h]
\caption{\label{tab:SRdefggF} Definition of the SRs used in the ggF selection. All SRs require at least two \LJs, but only the leading and the far \LJs\ are considered for the event classification. Dashes indicate cases where a requirement is not applied.}
\def\arraystretch{1.2}
\begin{center}
\begin{tabular}{l r r r}
\noalign{\smallskip}\hline\noalign{\smallskip}
Requirement / Region               & \SRggFmm & \SRggFhh & \SRggFhm \\
\noalign{\smallskip}\hline\noalign{\smallskip}
Number of \mLJs                    & 2        & 0        & 1        \\
Number of \hLJs                    & 0        & 2        & 1        \\
\hline
Tri-muon MS-only trigger           & yes      & -        & -        \\
Muon narrow-scan trigger           & yes      & -        & yes      \\
CalRatio trigger                   & -        & yes      & -        \\
\hline
$|\Delta t_{\textrm{\hLJs}}|$ [ns] & -        & $<2.5$   & -        \\
\hLJ JVT                           & -        & $<0.4$   & -        \\
$\Delta \phi_{\textrm{DPJ}}$       & $>\pi/5$ & $>\pi/5$ & $>\pi/5$ \\
BIB tagger score                   & -        & $>0.2$   & $>0.2$   \\
\maxiso[GeV]                      & $<4.5$   & $<4.5$   & $<4.5$   \\
\prodcalo                          & -        & $>0.95$  & $>0.9$   \\
\hline
\end{tabular}
\end{center}
\end{table}
 
The main sources of background after the SR selections are punch-through jets from rare multi-jet events for \SRggFmm\ and multi-jet production for \SRggFhh\ and \SRggFhm. The second leading background for all SRs is the cosmic-ray muon background.
 
In order to estimate the background contribution to each SR, a data-driven `ABCD' method is used. This method relies on the assumption that the  distribution of background events can be factorised in the plane of two uncorrelated variables so that it is subdivided into four regions: A, B, C, and D. The number of background events in the SR can be evaluated as $N_\mathrm{A} = N_\mathrm{D} \times N_\mathrm{B}/ N_\mathrm{C}$. Any possible signal leakage outside the SR region is accounted for by using a modified ABCD method that simultaneously fits the signal and background events, taking into account potential signal contamination in the ABCD estimate.
 
The planes used in the estimation of the backgrounds for the three ggF SRs are formed by \maxiso\ and either $\Delta \phi_{\textrm{DPJ}}$ in \SRggFmm, or \prodcalo\ in \SRggFhh and \SRggFhm. The definition of the three sets of ABCD control regions are given in Table~\ref{tab:CRdefggF}. In the ABCD convention, region A is the SR and regions B, C and D are respectively the control regions CRB, CRC and CRD.
 
\begin{table}[!h]
\caption{\label{tab:CRdefggF} Definition of the control regions used in the ggF background estimation. All control regions require at least two \LJs, but only the leading \LJ\ and the far \LJ\ are considered for the event classification. All CR requirements are the same as for the respective SR, with the exception of the selections reported in this table.}
\def\arraystretch{1.2}
\begin{center}
\begin{tabular}{l r r r}
\noalign{\smallskip}\hline\noalign{\smallskip}
Requirement / Region         & \CRBggFmm    & \CRCggFmm     & \CRDggFmm     \\
\noalign{\smallskip}\hline\noalign{\smallskip}
$\Delta \phi_{\textrm{DPJ}}$ & $>\pi/5$     & $(0.1,\pi/5]$ & $(0.1,\pi/5]$ \\
\maxiso[GeV]                & $[4.5 , 20)$ & $[4.5 , 20)$  & $<4.5$        \\
\noalign{\smallskip}\hline\noalign{\smallskip}
Requirement / Region         & \CRBggFhh    & \CRCggFhh     & \CRDggFhh     \\
\noalign{\smallskip}\hline\noalign{\smallskip}
\prodcalo                    & $>0.95$      & $(0.8, 0.95]$ & $(0.8, 0.95]$ \\
\maxiso[GeV]                & $[4.5 , 20)$ & $[4.5 , 20)$  & $<4.5$        \\
\noalign{\smallskip}\hline\noalign{\smallskip}
Requirement / Region         & \CRBggFhm    & \CRCggFhm     & \CRDggFhm     \\
\noalign{\smallskip}\hline\noalign{\smallskip}
\prodcalo                    & $>0.9$       & $(0.75, 0.9]$ & $(0.75, 0.9]$ \\
\maxiso[GeV]                & $[4.5 , 20)$ & $[4.5 , 20)$  & $<4.5$        \\
\hline
\end{tabular}
\end{center}
\end{table}
 
The cosmic-ray muon background contribution is estimated first, using the cosmic dataset. The ratio of the number of filled bunch crossings to the number of empty bunch crossings is used to scale the number of events in the cosmic dataset to that in the $pp$ collision data. The value of this correction depends on the trigger used in the SR selection and ranges from a minimum of 2.2 for the CalRatio trigger in 2015--2016 to a maximum of 3.9 for the tri-muon MS-only trigger in 2018. Only three regions were found to have a non-negligible cosmic-ray muon background contribution: $7\pm5$ events in \SRggFmm, $7\pm4$ events in \SRggFhh and $14\pm7$ events in \CRDggFhh. Since the events forming the cosmic dataset are collected in empty bunch crossings, there is no significant activity in the ID, which implies that the \sumpt is 0 for all reconstructed \LJs. For this reason, events populate only the SRs and CRD. These estimated cosmic-ray muon backgrounds are subtracted from the observed data in the main dataset before performing the ABCD procedure.
 
Events with BIB energy deposits are very likely to have two \hLJs which can contribute to the background. The possible contribution from misidentified \hLJs was studied using the BIB-enriched dataset. This selection is thus orthogonal to the signal regions.
The efficiency of the SR and CR selection requirements when applied to BIB-enriched events was found to be less than $2.3\cdot10^{-4}$, reducing the BIB contamination to a negligible contribution. The BIB background is hence neglected.
 
Figure~\ref{fig:ABCD_unblinded} shows the distribution of data events in the ABCD plane for the three search channels, together with the expected distribution for a benchmark FRVZ model assuming a 125~\GeV\ Higgs boson and a decay branching fraction to the dark sector of 10\%.
The acceptance times efficiency for the FRVZ signal processes after applying all selection criteria is 0.74\%, 0.02\% and 0.15\%, in the case of a \darky mass of 400~\MeV\ and a generated \ctau of 50~mm, for the signal regions \SRggFmm, \SRggFhh and \SRggFhm respectively.
 
\begin{figure}[h]
\centering
\subfloat[\SRggFmm]{\includegraphics[width=0.3\textwidth]{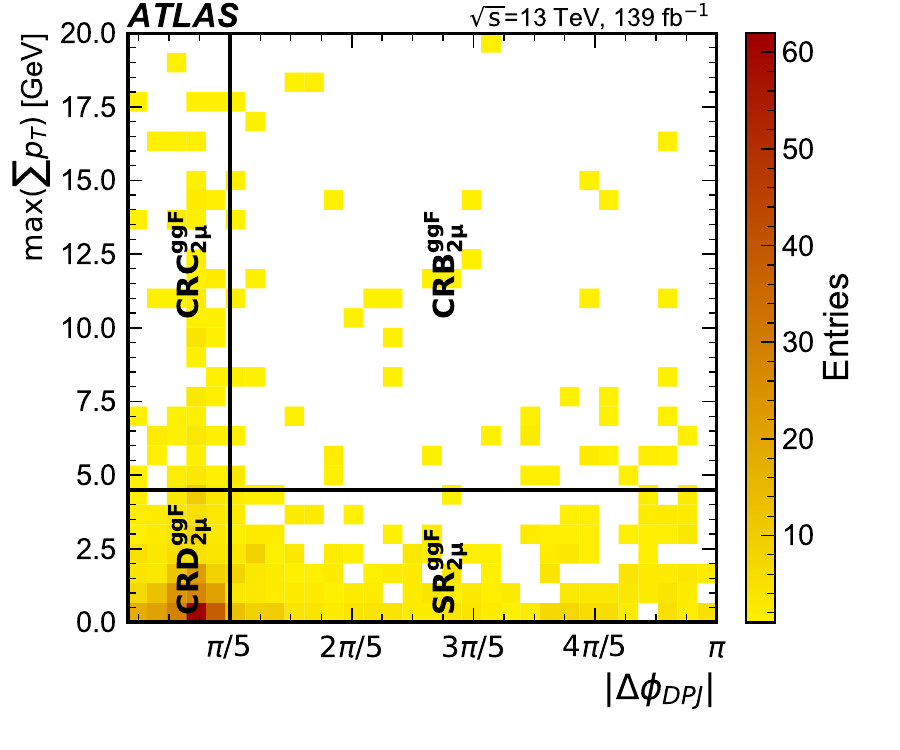}} 
\subfloat[\SRggFhm]{\includegraphics[width=0.3\textwidth]{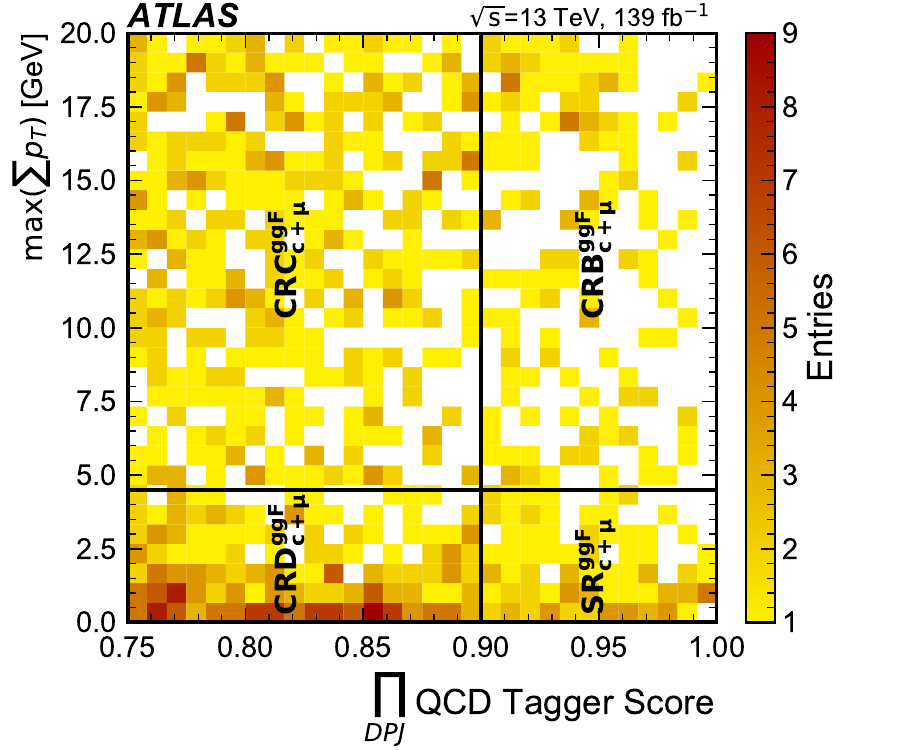}}  
\subfloat[\SRggFhh]{\includegraphics[width=0.3\textwidth]{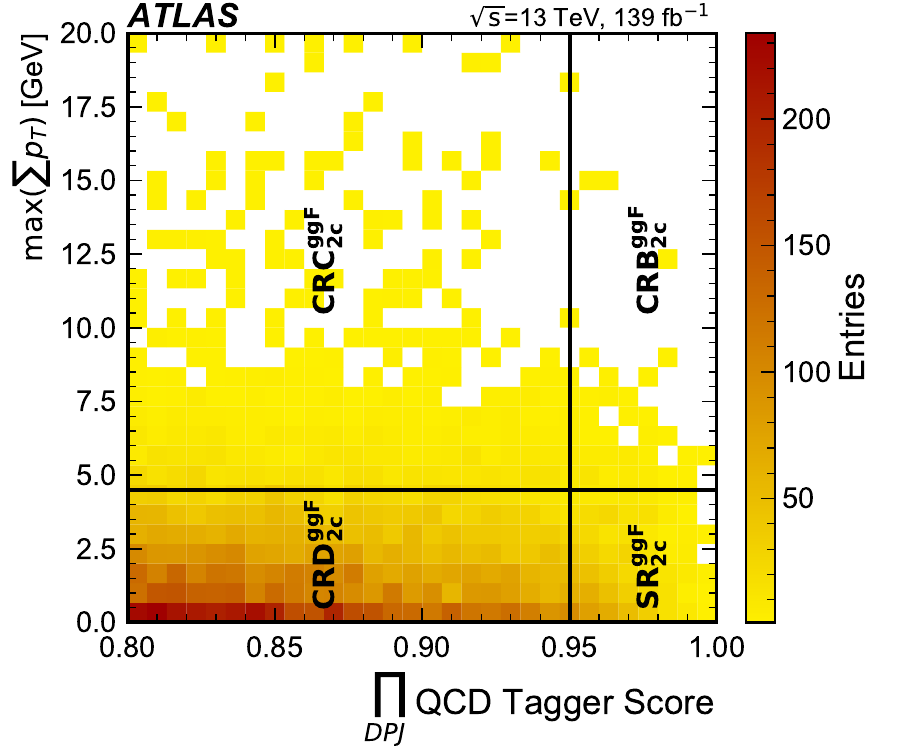}}\\
\subfloat[\SRggFmm]{\includegraphics[width=0.3\textwidth]{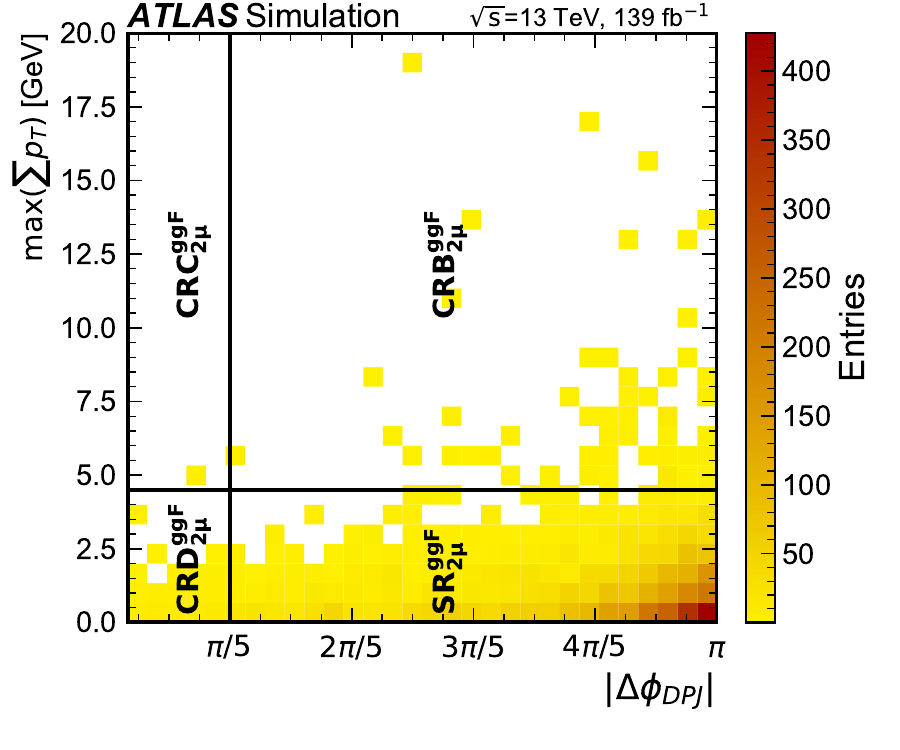}}
\subfloat[\SRggFhm]{\includegraphics[width=0.3\textwidth]{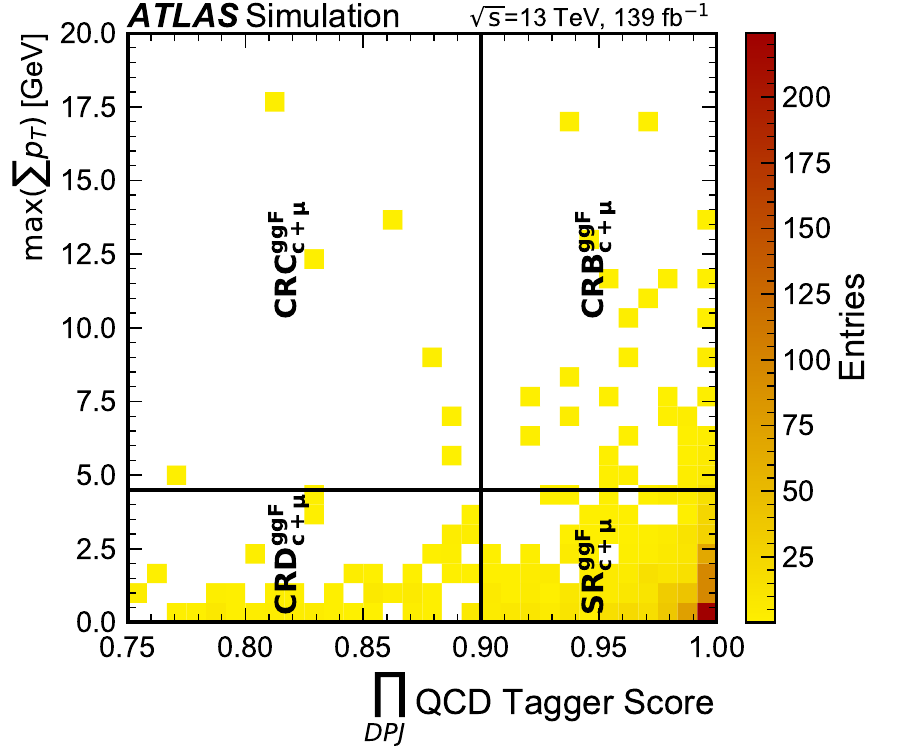}}
\subfloat[\SRggFhh]{\includegraphics[width=0.3\textwidth]{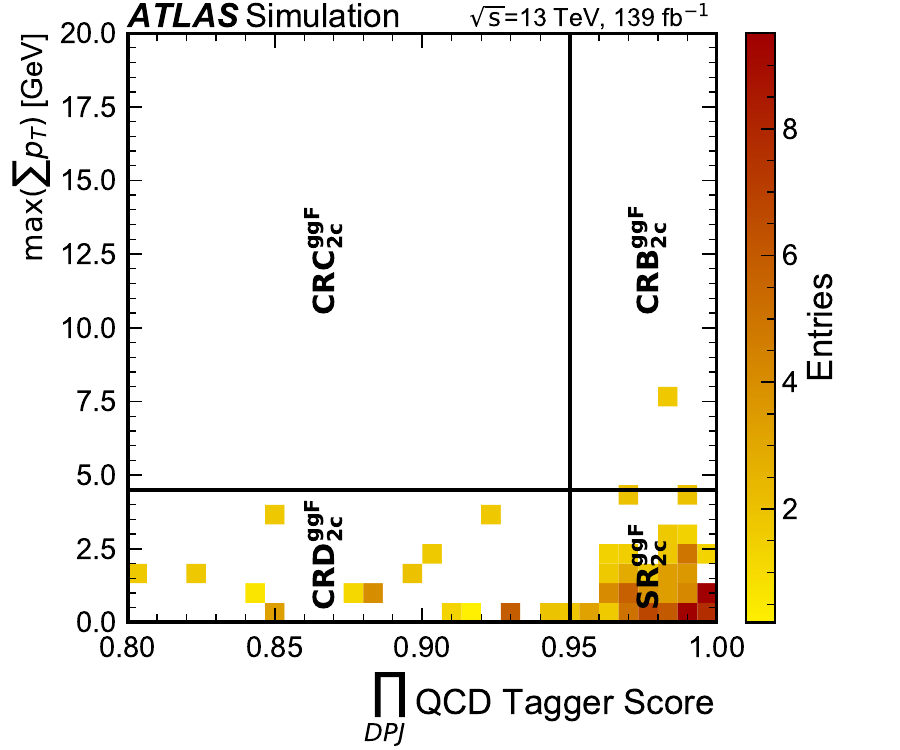}}
\caption{The per-event distributions in the ABCD planes defined for the ggF search channels. Figures~(a, b, c) show data events, while Figures~(d, e, f) show simulated signal events. FRVZ signal samples assuming a 125~\GeV\ Higgs boson with a decay branching fraction of 10\% to the dark sector and a \darky mass of 400~\MeV\ are shown.}
\label{fig:ABCD_unblinded}
\end{figure}
 
The background estimation procedure was validated by applying the ABCD method in subregions constructed in combinations of the regions CRB, CRC and CRD, for varying choices of the subregion boundaries. For each validation test, a new set of (ABCD)$'$ test regions was defined within the union of either CRB and CRC or CRC and CRD. The test region boundaries were varied in discrete steps after dividing the allowed range into ten equal-size intervals. The validation was performed for each of the \SRggFmm, \SRggFhh\ and \SRggFhm selections.
The expected background yields in the resulting test SRs range between 15 and 800 events.
An additional validation of the method was performed by checking the closure on orthogonal selections, which were obtained either by inverting the \mLJtagger\ cut or by selecting different ranges of \hLJtagger\ score. These selections were chosen in order to have negligible signal contributions. In all the channels, the observed yields were found to agree within one standard deviation with the expected yields. The Pearson linear correlation coefficient between the two variables defining the ABCD plane in all subregions and in the validation regions was found to be less than 3\%.
The signal leakage in regions CRB, CRC and CRD, was found to be less than 10\% of the total signal in the ABCD selection for all signal scenarios considered in the analysis.


\FloatBarrier

\subsection{\textit{WH} associated production selection}
\label{sec:Wh}
 
In the \textit{WH} selection, candidate events are required to pass single-electron or single-muon triggers. The offline requirements on the \pT, identification and isolation of the lepton are tighter than those applied online, so as to be on the trigger efficiency plateau~\cite{TRIG-2016-01}.
 
The selection requires one electron or muon with \pt~$\ge$~27~\GeV, \MET~$\ge$~30~\GeV, $m_\mathrm{T}~\ge$~40~\GeV, and no additional lepton with \pt~$\ge$~10~\GeV.
Events where the two leading jets have an invariant mass $m_{\textrm{jj}}\ge 1$~\TeV\ are removed to allow for a future statistically independent study of VBF production modes. Contributions from single-top and $t\bar{t}$ processes are reduced by requiring three or fewer jets with \pT~$\ge$~30~\GeV\ and no $b$-tagged jets in each event.
After these selections the background is dominated by \textit{W}+jets events.
 
Events are then separated into orthogonal search channels based on the numbers of \mLJs and \hLJs. Events with exactly one \hLJ are targeted by \SRWHh, and events with two or more \hLJs by \SRWHhh. The mixed channel with one \LJ of each type is targeted by \SRWHhm. Channels with only \mLJs are not considered for the \textit{WH} selection as the constraints they provide are not competitive with those from other signal production modes. The SR requirements are summarised in Table~\ref{tab:SRdefWH}.
In all SRs, \hLJ candidates are required to have a JVT score below 0.6 to suppress jets originating from the interaction point.
Events belonging to \SRWHh, where one \darky is reconstructed, are required to have $m_\mathrm{T}~\ge$~120~\GeV\ and a \hLJ with \pt~$\ge$~30~\GeV\ to further reduce the initially overwhelming \textit{W}+jets background. In this SR, the second \darky is expected to decay outside the detector volume and yield a higher \MET. CaloDPJ candidates are required to have a width $w\le$~0.08 in \SRWHh and $w\le$~0.1 in \SRWHhm. In \SRWHhh\ the leading (far) \hLJ candidate is required to have $w\le$~0.1 (0.15). These selections are optimised to reject hadronic jets and maximise the discovery significance.
\begin{table}[!h]
\caption{\label{tab:SRdefWH} Definition of the signal regions used in the \textit{WH} selection. In signal regions requiring at least two \LJs, only the leading \LJ\ and the far \LJ\ are considered for the event classification. Each \LJ SR is exclusive in the number of \LJs in the event. Dashes indicate cases where a requirement is not applied.}
\def\arraystretch{1.2}
\begin{center}
\begin{tabular}{l r r r}
\noalign{\smallskip}\hline\noalign{\smallskip}
Requirement / Region               & \SRWHh    & \SRWHhh         & \SRWHhm    \\
\noalign{\smallskip}\hline\noalign{\smallskip}
Number of \mLJs                    & 0         & 0               & 1          \\
Number of \hLJs                    & 1         & 2               & 1          \\
\hline
Single-lepton trigger ($\mu$, $e$) & yes       & yes             & yes        \\
\hline
$m_\mathrm{T}$ [\GeV]              & $>120$    & -               & -          \\
$|t_{\mathrm{\hLJ}}|$ [ns]         & $<4$      & $<4$            & $<4$       \\
Leading (far) \hLJ width           & $<0.08$   & $<0.10\ (0.15)$ & $<0.1$     \\
\hLJ $p_\mathrm{T}$ [\GeV]         & $>30$     & -               & -          \\
JVT                                & $<0.6$    & $<0.6$          & $<0.6$     \\
\mindphi                           & $<3\pi/5$ & $<3\pi/10$      & $<7\pi/20$ \\
\minhtag                           & $>0.99$   & $>0.91$         & $>0.9$     \\
\hline
\end{tabular}
\end{center}
\end{table}
 
After the selection the main background is due to \textit{W}+jets events where rare QCD phenomena give rise to in-flight decays and punch-through jets that mimic signal \LJs.
The residual \textit{W}+jets background in the signal region is estimated with a data-driven ABCD method, the same as used for the ggF selection and explained in detail in Section~\ref{sec:ggF}. The non-collision background from cosmic-ray muons and BIB was found to be negligible for this selection.
 
The ABCD planes used in the estimation of the backgrounds for the three \textit{WH} SRs are based on \mindphi and \minhtag. These variables provide good discrimination between signal and background, and are highly uncorrelated in \textit{W}+jets events. The definitions of the three sets of ABCD regions are given in Table~\ref{tab:CRdefWH}.
 
\begin{table}[!h]
\caption{\label{tab:CRdefWH} Definition of the control regions used in the \textit{WH} background estimation. The requirements for all regions are the same as those for the respective SRs, except for the selections reported in this table.}
\def\arraystretch{1.2}
\begin{center}
\begin{tabular}{l r r r}
\noalign{\smallskip}\hline\noalign{\smallskip}
Requirement / Region & \CRBWHh        & \CRCWHh        & \CRDWHh    \\
\noalign{\smallskip}\hline\noalign{\smallskip}
\mindphi             & $<3\pi/5$      & $>3\pi/5$      & $>3\pi/5$  \\
\minhtag             & $[0.9 , 0.99)$ & $[0.9 , 0.99)$ & $>0.99$    \\
\noalign{\smallskip}\hline\noalign{\smallskip}
Requirement / Region & \CRBWHhh       & \CRCWHhh       & \CRDWHhh   \\
\noalign{\smallskip}\hline\noalign{\smallskip}
\mindphi             & $<3\pi/10$     & $>3\pi/10$     & $>3\pi/10$ \\
\minhtag             & $[0.8 , 0.91)$ & $[0.8 , 0.91)$ & $>0.91$    \\
\noalign{\smallskip}\hline\noalign{\smallskip}
Requirement / Region & \CRBWHhm       & \CRCWHhm       & \CRDWHhm   \\
\noalign{\smallskip}\hline\noalign{\smallskip}
\mindphi             & $<7\pi/20$     & $>7\pi/20$     & $>7\pi/20$ \\
\minhtag             & $[0.8 , 0.9)$  & $[0.8 , 0.9)$  & $>0.9$     \\
\hline
\end{tabular}
\end{center}
\end{table}
All the SRs lie at high \hLJtagger score and small azimuthal angular separation between the \ptmiss and the nearest \LJ, where most of the signal events are expected.
 
The background estimation procedure was validated by applying the ABCD method in subregions constructed in combinations of the regions CRB, CRC and CRD, for various choices of the subregion boundaries, as done for the ggF selection. The expected background yields in the resulting test SRs range between 80 and 1550 events. Agreement within one standard deviation between data and the predicted yields was observed in most regions, with discrepancies of up to two standard deviations appearing in the statistically limited validation region for \SRWHh for values of \minhtag close to 0.9.
An additional validation of the method was performed by checking the closure of the ABCD procedure on orthogonal selections, which were obtained by selecting different ranges of QCD tagger score. The validation ranges extend to a minimum of 0.6. In all the channels, the observed yields were found to agree within one standard deviation with the expected yields.
The Pearson linear correlation coefficient between the two variables defining the ABCD plane in all subregions and in the validation regions was found to be less than 2\%.
 
Figure~\ref{fig:wh_ABCD_unblinded} shows the distribution of data events in the ABCD plane for the three search channels, together with the expected distribution for a benchmark FRVZ model assuming a 125~\GeV\ Higgs boson.
The acceptance times efficiency for the FRVZ signal processes after applying all selection criteria is 0.12\% in the case of a \darky mass of 400~\MeV\ and \ctau of 50~mm in \SRWHhm, and 0.05\% and 0.26\% in \SRWHhh and \SRWHh, respectively, in the case of a \darky mass of 100~\MeV\ and \ctau of 15~mm.
The signal leakage in the CRs is accounted for in the fit and can be as high as 25\% for CRB, and up to 15\% for CRD and CRC. However, the stability of the fit is not affected by the potential signal leakage because the expected signal yields are negligible in comparison with the SM background in those regions.

\begin{figure}[h]
\centering
\subfloat[\SRWHh]{\includegraphics[width=0.30\textwidth]{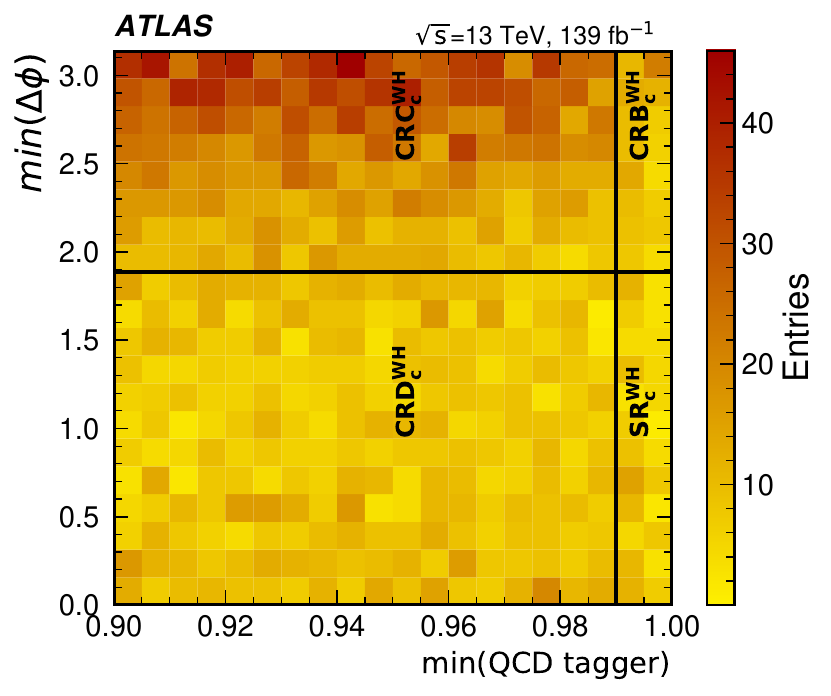}}
\subfloat[\SRWHhh]{\includegraphics[width=0.30\textwidth]{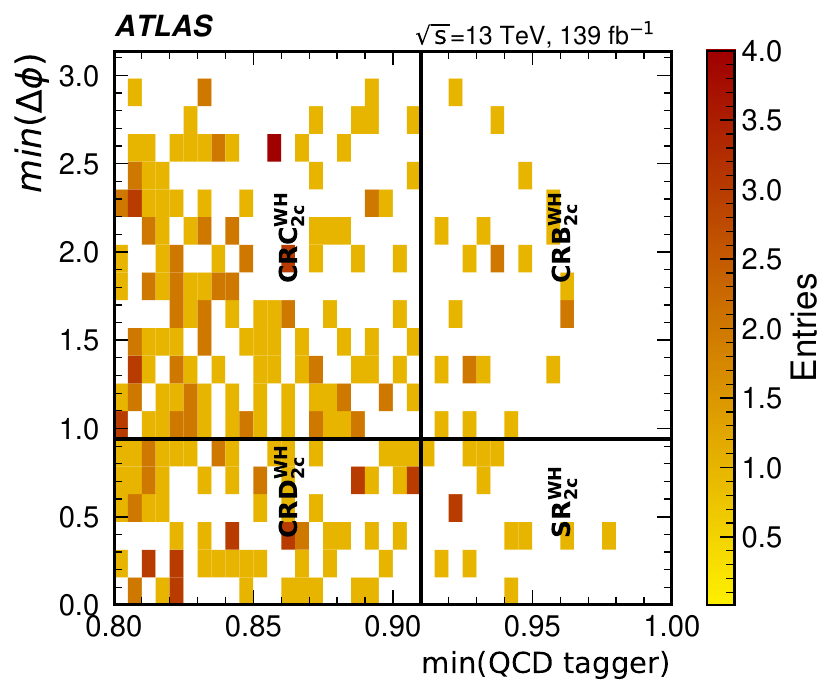}}
\subfloat[\SRWHhm]{\includegraphics[width=0.30\textwidth]{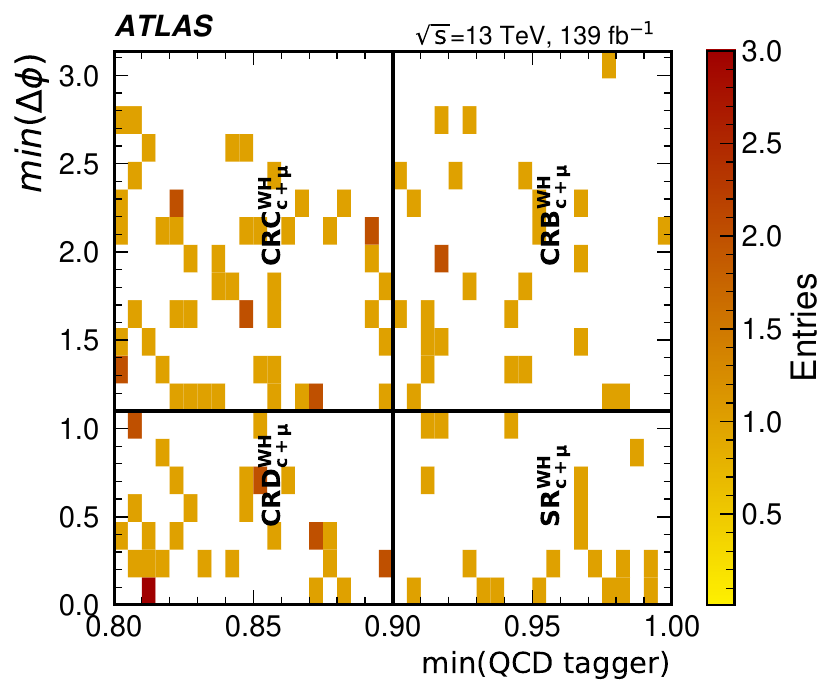}}\\
\subfloat[\SRWHh]{\includegraphics[width=0.3\textwidth]{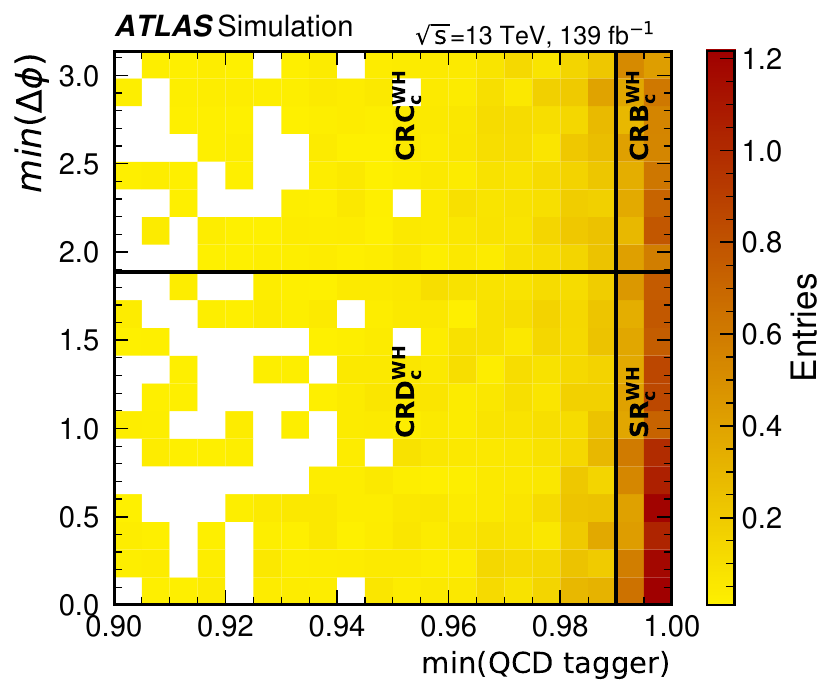}}
\subfloat[\SRWHhh]{\includegraphics[width=0.3\textwidth]{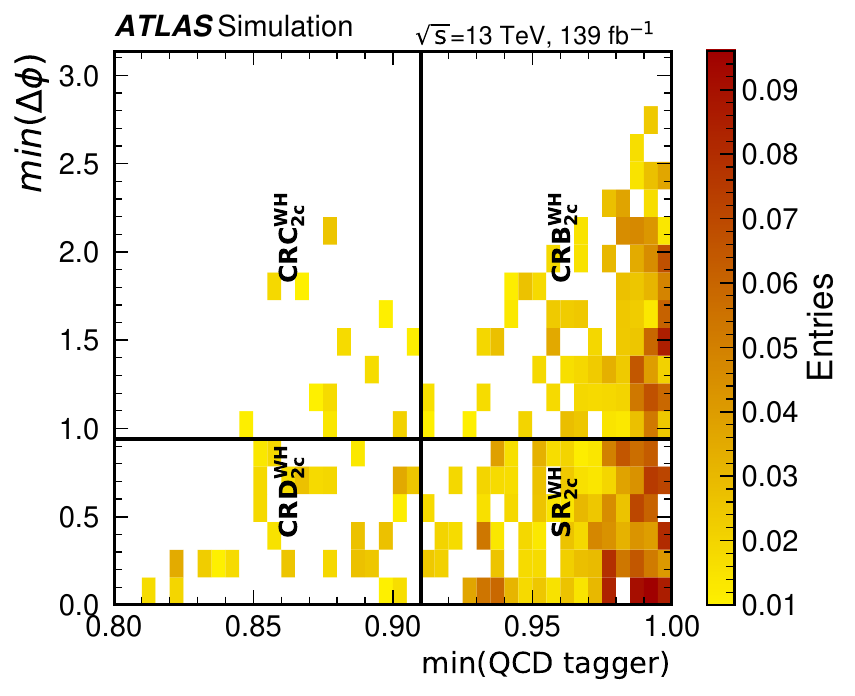}}
\subfloat[\SRWHhm]{\includegraphics[width=0.3\textwidth]{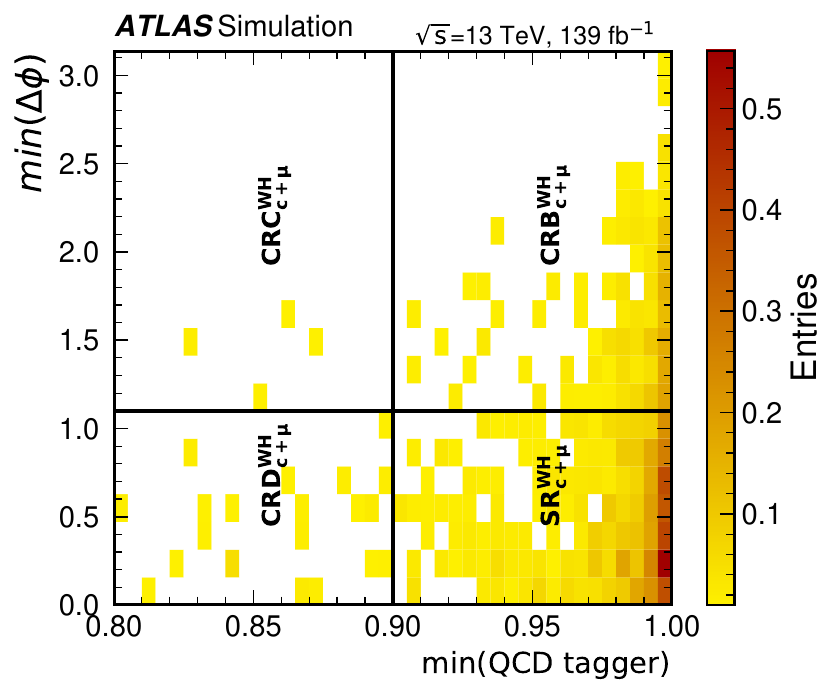}}
\caption{The per-event \mindphi vs \minhtag distributions for the \textit{WH} search channels. Figures~(a, b, c) show data, while Figures~(d, e, f) show simulated signal events. FRVZ signal samples with a 125~\GeV\ Higgs boson and a \darky mass of 100~\MeV\ are shown for \SRWHh and \SRWHhh, while a \darky mass of 400~\MeV\ is shown for \SRWHhm.}
\label{fig:wh_ABCD_unblinded}
\end{figure}


\FloatBarrier

\section{Systematic uncertainties}
\label{sec:uncertainties}
The overall uncertainty in the SR yields is dominated by the statistical uncertainty. Nevertheless, potential sources of experimental uncertainty are considered for the background estimates and the simulated signal yields.
 
The statistical uncertainties of the observed yields in regions CRB, CRC, and CRD are propagated to the background expectation obtained from the ABCD method. An additional uncertainty is assigned to account for the size of the cosmic dataset and scaled by the ratio of the number of empty bunch crossings to the number of filled bunch crossings while the triggers were active. This uncertainty is estimated to be as large as 80\% of the expected cosmic-ray muon background, but has no impact on the final results because of the limited contribution from this specific background.
 
Experimental uncertainties in the reconstruction and simulation of the signal events are considered. For jets, these include jet energy resolution (JER) and jet energy scale (JES) uncertainties from the standard calibration scheme~\cite{PERF-2016-04}, which amount to up to 3\% of the expected signal yields. An extra JES uncertainty is applied to take into account a possible dependence on the low-EM-fraction selection. It is estimated following the same procedure as used in the 2015--2016 dark-photon jet search~\cite{EXOT-2017-28} and is about 3\%.
 
For muons, a systematic uncertainty in the single-\darky reconstruction efficiency is evaluated using a tag-and-probe method applied to $J/\psi \to \mu\mu$ events in data and simulation, as $J/\psi \to \mu\mu$ decays were found to behave similarly to \darky\ decays in simulated events. The  $J/\psi$ reconstruction efficiency is evaluated in both data and simulation as a function of the opening angle \delr between the two muons. For low \delr values, the efficiency decreases due to the difficulty of reconstructing two tracks with small angular separation in the MS. The difference in $J/\psi \to \mu\mu$ reconstruction efficiency between simulation and data in the \delr~interval between 0 and 0.06, where the DPJ samples are concentrated, amounts to 9.6$\%$ and is taken as the uncertainty. This uncertainty includes a subdominant contribution of approximately 0.1\% on the matching of MS and ID tracks.
 
The systematic uncertainties in the efficiency of the two dedicated muon triggers are evaluated using a tag-and-probe method applied to $J/\psi \to \mu\mu$ events in data and simulation. The difference between the trigger efficiency in data and that in simulation is evaluated as a function of the opening angle between the two muons. The difference in the region \delr~<~0.05, corresponding to the \delr expected for signal, is taken as the uncertainty and is found to be 6$\%$ for the muon narrow-scan trigger and 5.8$\%$ for the tri-muon MS-only trigger.
 
The systematic uncertainty in the CalRatio trigger efficiency is taken from Ref.~\cite{EXOT-2019-23} and recomputed for each signal sample. It is estimated to be up to 4\% for the heavy (800~\GeV) scalar-mediator signal samples, and up to 30\% for the 125~\GeV\ Higgs signal samples.
 
The uncertainty associated with the MC modelling of the input variables used for the cosmic-ray tagger training is evaluated by comparing the distributions of the tagger's output scores for $Z\rightarrow\mu\mu$ events in data and MC samples.
Due to the high boost of low-mass dark photons and the high $p_\mathrm{T}$ of signal muons, the muon $z_0$ and timing distributions are similar to those of prompt muons from $Z\rightarrow\mu\mu$ decays. Muons from $Z$ boson decays were reconstructed using only MS information.
The ratio of the two score distributions is computed and applied as a binned scale factor to the muons used for \mLJ reconstruction.
The resulting relative variation of the final signal yield in the SR is taken as the systematic uncertainty.
The same procedure is used for the QCD and BIB taggers, where the ratios of data to simulated distributions are computed from data and MC samples of multi-jet events after selecting signal-like events with reconstructed \hLJs\ but failing the analysis trigger selections. In-depth modelling studies of jet calorimeter-cell clusters, upon which the taggers' training images are built, show no significant mis-modelling.
 
A pile-up modelling uncertainty is assigned to account for the difference between the predicted and measured inelastic cross-sections~\cite{STDM-2015-05}.
 
Finally, an uncertainty in the computed total integrated luminosity used to rescale the expected number of signal events is considered. The uncertainty in the combined 2015--2018 integrated luminosity is 1.7\%~\cite{ATLAS-CONF-2019-021}, obtained using the LUCID-2 detector~\cite{LUCID2} for the primary luminosity measurements.
 
A summary of the experimental systematic uncertainties taken into account for the signal samples in this analysis is shown in Figure~\ref{fig:sysPlot}, assuming signal production via a 125~\GeV\ Higgs boson. The average uncertainties are representative for the bulk of the signal samples, with small variations of the order of a few percent being observed as a function of the \darky\ mass. The uncertainty related to the efficiency of the CalRatio trigger was found to have the largest variation, ranging between 15\% and 26\% with a linear dependence on the \darky\ mass.
\begin{figure}[ht!]
\centering
\includegraphics[width=.95\textwidth]{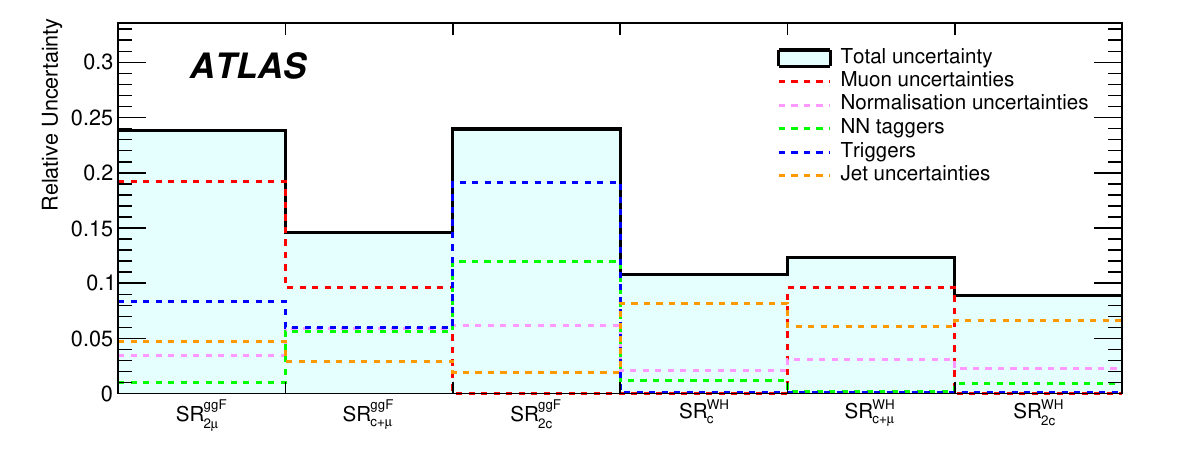}
\caption{
Comparison of the relative uncertainty in the signal yield in each SR, showing the contributions from the different sources of uncertainty. The quoted values are averaged over different \darky\ masses, in scenarios assuming signal production via a 125~\GeV\ Higgs boson. The `Muon uncertainties' category contains all muon-related systematic uncertainties and is dominated by the uncertainty in the single-\darky\ reconstruction efficiency. The `NN taggers' category contains the three taggers adopted in the analysis and is dominated by the BIB tagger uncertainties for the hadronic channels and by the cosmic-ray tagger in the muonic channels. The `Triggers' category contains all trigger systematic uncertainties, which are relevant only for the dedicated triggers in the ggF selection. The `Jet uncertainties' category contains the JES, JER and low-EM-fraction JES systematic uncertainties. Some sets of systematic uncertainties apply to only a subset of the SRs.
}
\label{fig:sysPlot}
\end{figure}

\FloatBarrier


\FloatBarrier

\section{Results and interpretations}
\label{sec:result}
The data-driven background estimate in each SR is obtained by performing a maximum-likelihood fit to the yields in the four  (i.e.\ ABCD) regions in data.
The fitted likelihood function is formed from a product of Poisson functions, one for each of the SR, CRB, CRC, and CRD regions, describing signal and background expectations. The ABCD ansatz is introduced as nuisance parameters in the background component of the expected yield in each region.
The likelihood-based ABCD fit is robust against control regions with a small number of events and takes into account possible signal contamination in the control regions.
All systematic uncertainties described in Section~\ref{sec:uncertainties} are included in the fit as nuisance parameters, parameterised with Gaussian probability density functions that multiply the fit likelihood. They are assumed to be uncorrelated across regions.
An alternative correlation model, where the uncertainties are assumed to be fully correlated across regions, has a negligible impact on the results.
The mean value of the Gaussian probability distribution function is constrained by the nominal value of the parameter and the variance is defined by the 68\% confidence interval of the systematic uncertainty associated with the parameter.
 
The observed and expected numbers of events in the signal regions are summarised in Table~\ref{tab:DataAbcd}.
The reported yields are extrapolated by the fit assuming no signal, and with unblinded data in all ABCD regions in the fit.
When comparing the results with a likelihood fit using blinded data in the SR, the background yields are found to be consistent at the percent level except for \SRggFmm, where an 11\% change is observed with a relative uncertainty reduction by 32\%.
For the ggF selection results, the estimated cosmic-ray muon yields are subtracted from each of the ABCD regions before using the ABCD method to estimate the multi-jet background yield.
 
\begin{table}[h]
\caption{Observed and expected yields in the ABCD regions. The total uncertainty in the background expectation is computed by the ABCD fit to unblinded data. In the ggF selection regions the estimated contribution from cosmic-ray muons is subtracted from each of the ABCD regions before the ABCD fit, and added back post-fit (as shown in this table).}
\label{tab:DataAbcd}
\def\arraystretch{1.2}
\centering
\begin{tabular}{ccccccc}
\hline
Selection   & Search channel & CRB    & CRC    & CRD      & SR expected     & SR observed \\  \hline
\hline
& $2\mu$         & ~~~~55 & ~~~~61 & ~~~~389  & ~~$317 \pm 47$  & ~~269       \\
ggF         & c+$\mu$        & ~~169  & ~~471  & ~~~~301  & ~~$108 \pm 13$  & ~~110       \\
& 2c             & ~~~~97 & 1113   & 12146    & $1055 \pm 82$   & 1045        \\
\hline
& c              & 1850   & 3011   & ~~~~155  & ~~~~$93 \pm 12$ & ~~103       \\
\textit{WH} & c+$\mu$        & ~~~~30 & ~~~~49 & ~~~~~~31 & ~~$19 \pm 8$    & ~~~~20      \\
& 2c             & ~~~~79 & ~~155  & ~~~~~~27 & ~~$14 \pm 5$    & ~~~~15      \\
\hline
\hline
\end{tabular}
\end{table}
 
The results of the search are used to set upper limits on the production cross-section times branching fraction, \sigmabr, of the process \gghHtwogd, with $X$ assumed to be undetected, as a function of the \darky mean proper decay length \ctau. The upper limit on the signal strength is obtained with the CL$_\text{s}$ method~\cite{Read:2002hq} with the asymptotic calculator~\cite{Cowan:2010js}, considering the background and the predicted signal yields from simulation in the four ABCD regions. The validity of the asymptotic approximation is checked against a full calculation using pseudo-experiments, and the CL$_\text{s}$ values of the two methods typically agree within 1\%, and maximally within 5\%.
Model-dependent limits are computed for the various signal scenarios considered in the analysis. The hypothesis tests take account of the expected signal yield and its uncertainties in the CRs and SRs and are performed with a likelihood fit to the observed data.
 
A summary of the observed limits for scenarios assuming the production of a 125~\GeV\ Higgs boson that can decay into hidden-sector particles, resulting in two \darky\ in an FRVZ model, is shown in Figures~\ref{fig:limits_allggF} and~\ref{fig:limits_allWH} for the ggF and \textit{WH} selections, respectively.
The sensitivity of each search channel is reported in a separate subfigure. The sensitivity of the $2\mu$ and c+$\mu$ selections to \darky\ masses less than twice the muon mass is due to signal decays into a displaced electron or quark pair within the muon spectrometer volume which are reconstructed as \mLJs.
For pairs of dark photons decaying into \mLJs, branching fractions larger than 1\% are excluded at 95\% confidence level (CL) if the dark photons have a mean proper decay length \ctau between 10~mm and 250~mm and a mass between 0.4~\GeV\ and 2~\GeV. When considering dark-photon decays into pairs of \hLJs, branching fractions larger than 10\% are excluded at 95\% CL for mean proper decay lengths between 2~mm and 3~mm and masses between 17~\MeV\ and 50~\MeV.
 
\begin{figure}[h!]
\centering
\subfloat[\SRggFmm]{\includegraphics[width=.49\textwidth]{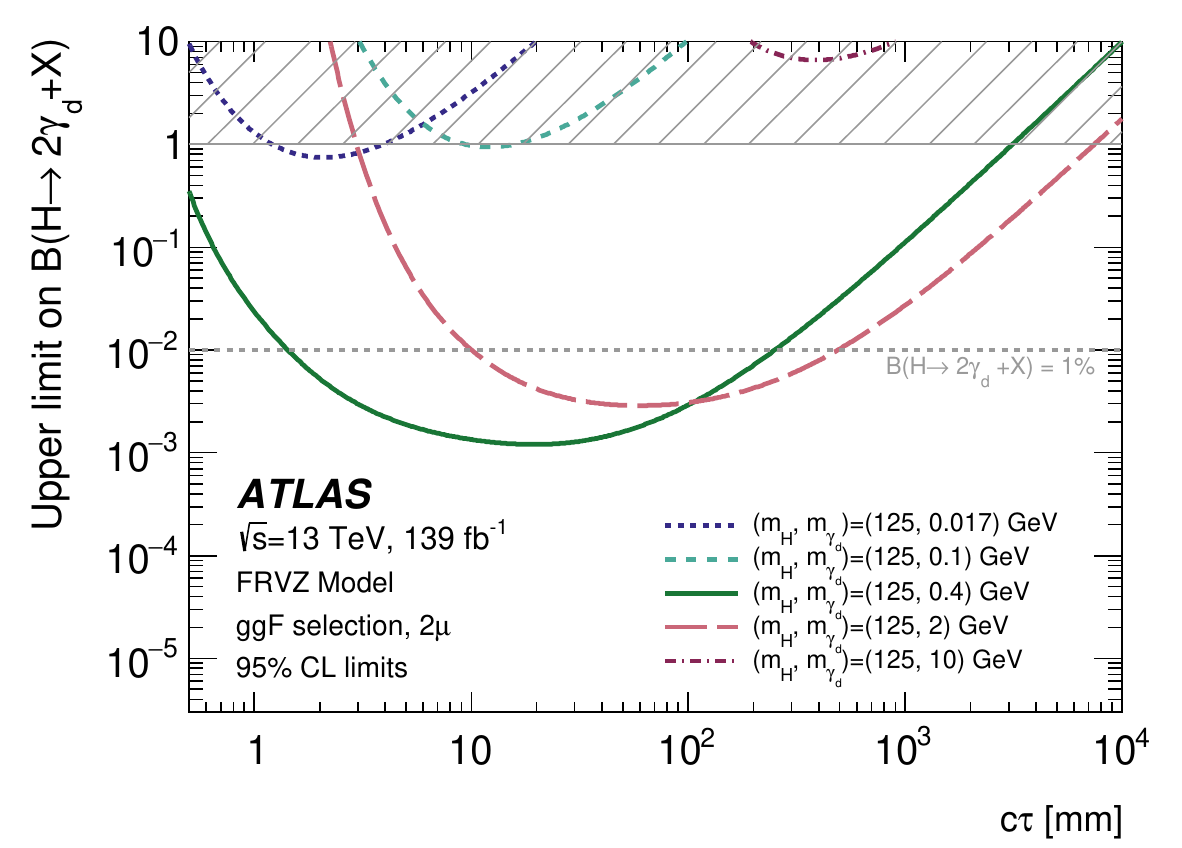}}
\subfloat[\SRggFhm]{\includegraphics[width=.49\textwidth]{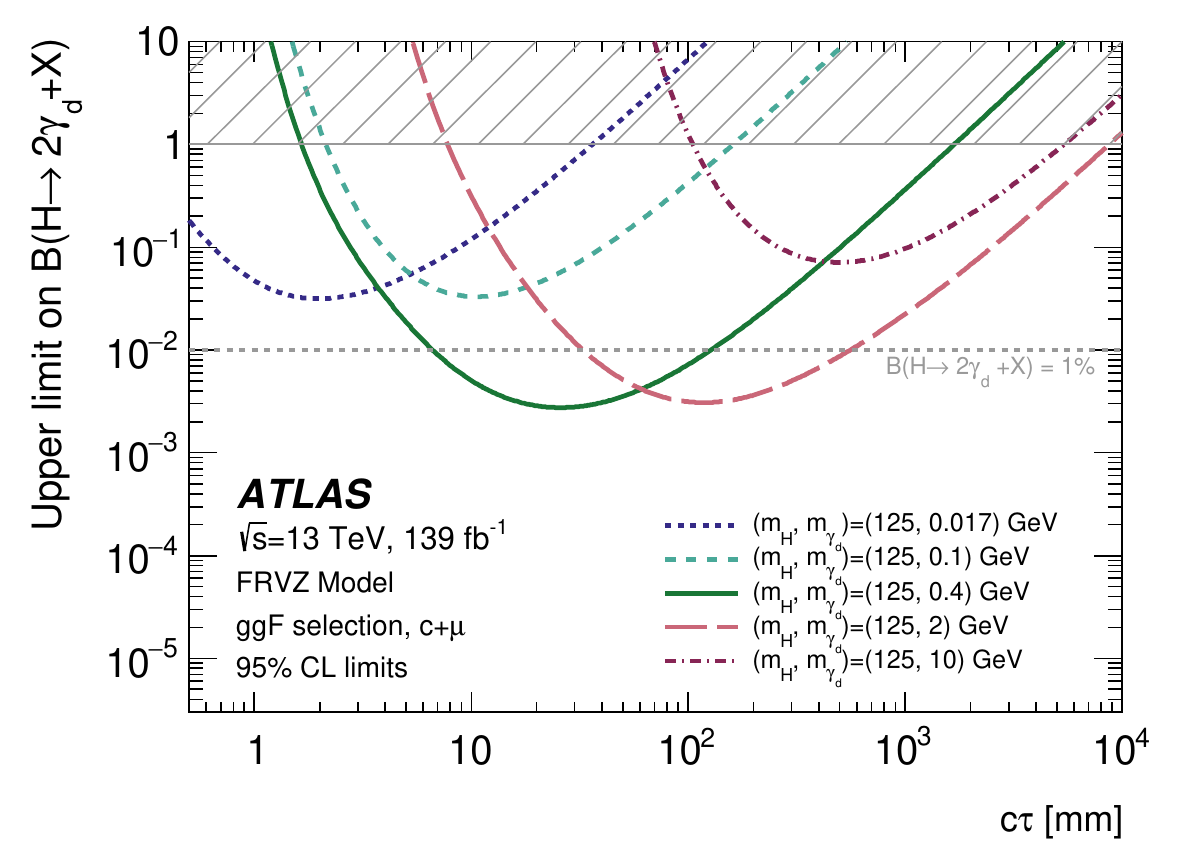}}\\
\subfloat[\SRggFhh]{\includegraphics[width=.49\textwidth]{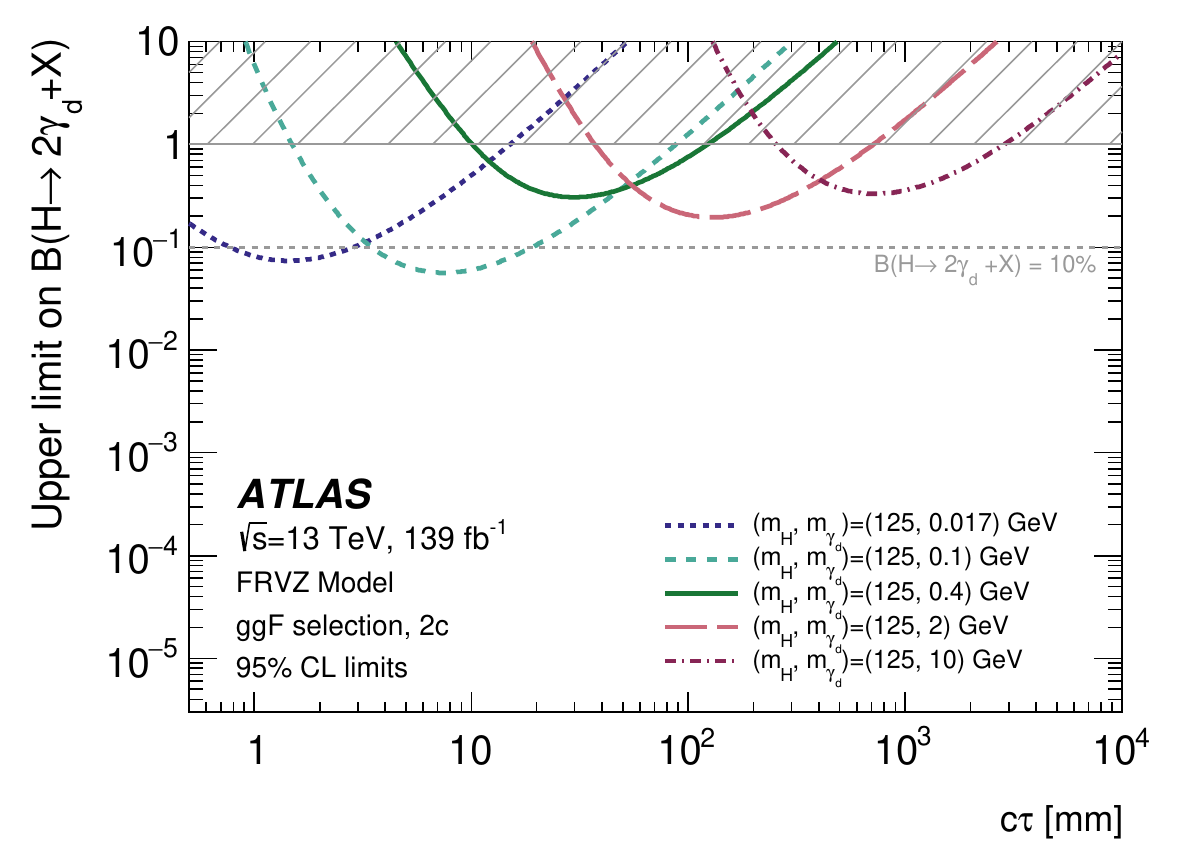}}
\caption{Observed 95$\%$ CL upper limits on the branching ratio ($B$) for the decay \Htwogd and ggF selection, for different \darky\ masses and a 125~\GeV\ Higgs boson. The limits are shown separately for the (a)~$2\mu$, (b)~c$+\mu$ and (c)~2c search channels, assuming an FRVZ signal model. The hatched band denotes the region in which the branching ratio of \Htwogd is larger than unity.}
\label{fig:limits_allggF}
\end{figure}
 
\begin{figure}[h!]
\centering
\subfloat[\SRWHh]{\includegraphics[width=.49\textwidth]{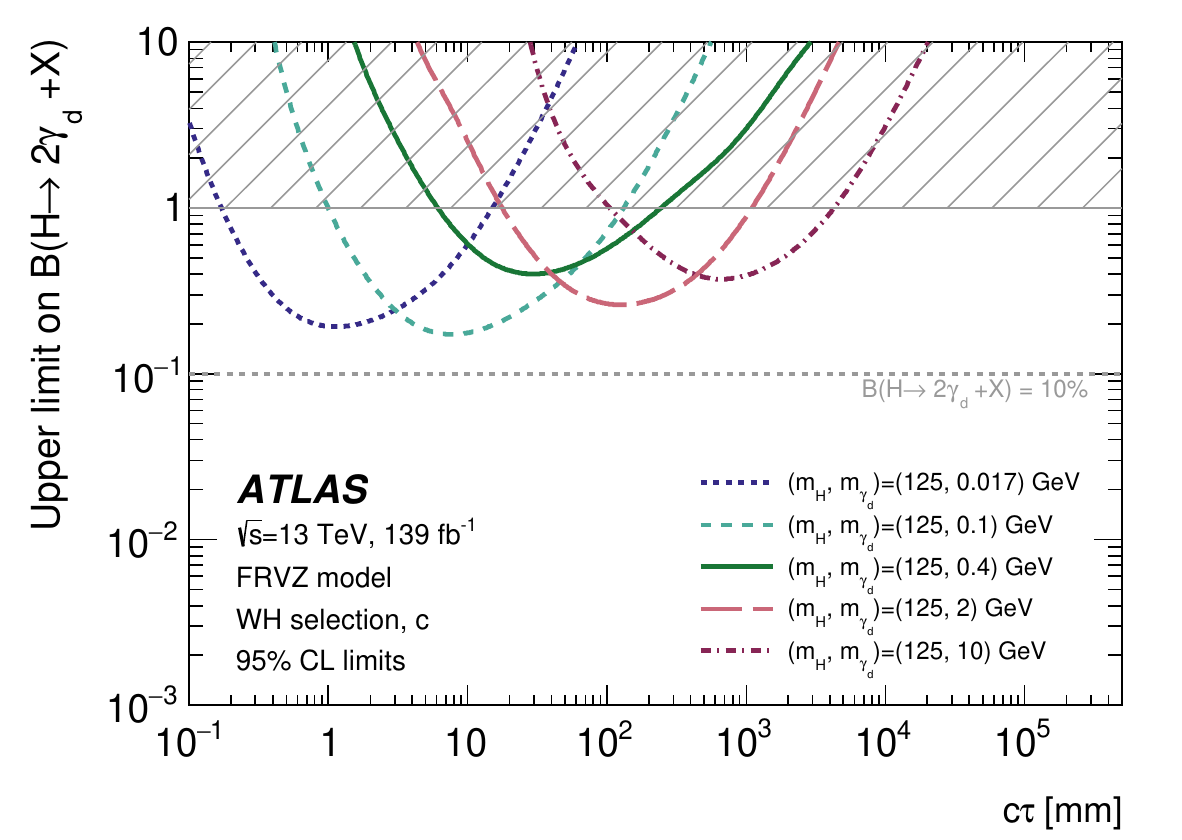}}
\subfloat[\SRWHhm]{\includegraphics[width=.49\textwidth]{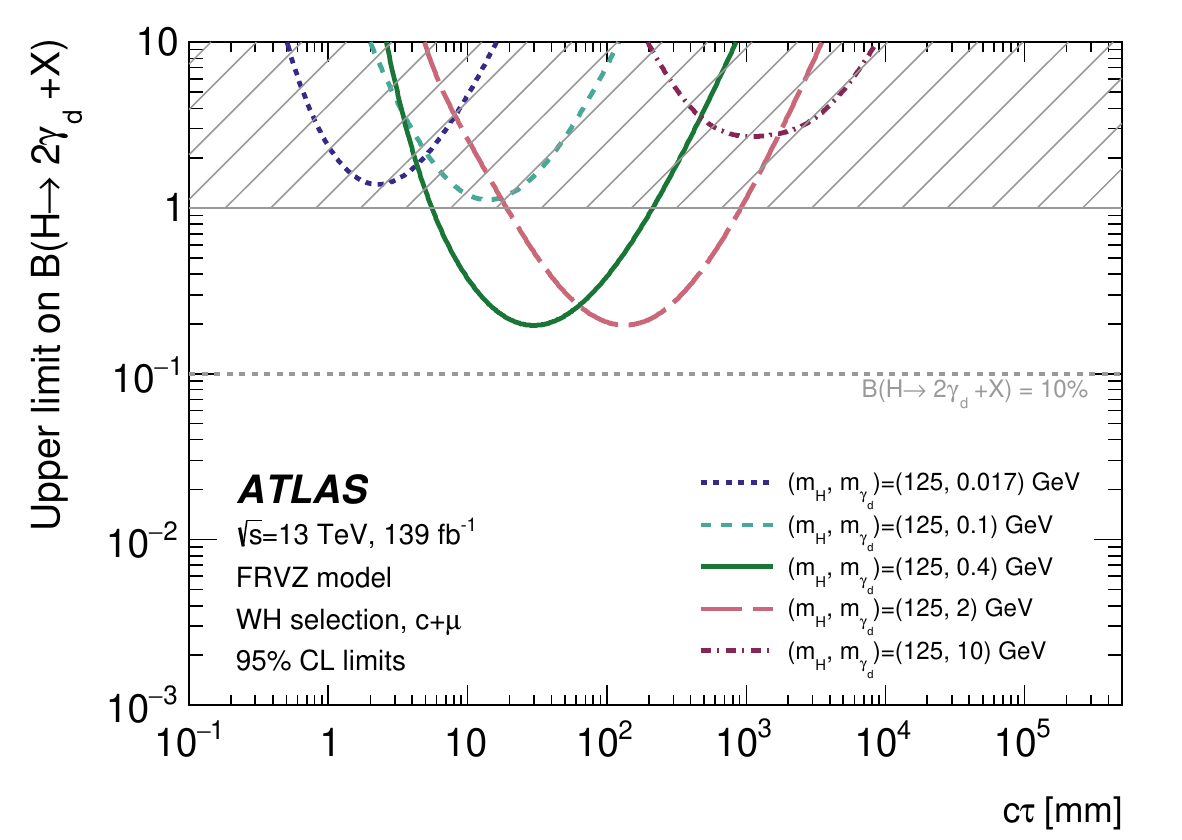}}\\
\subfloat[\SRWHhh]{\includegraphics[width=.49\textwidth]{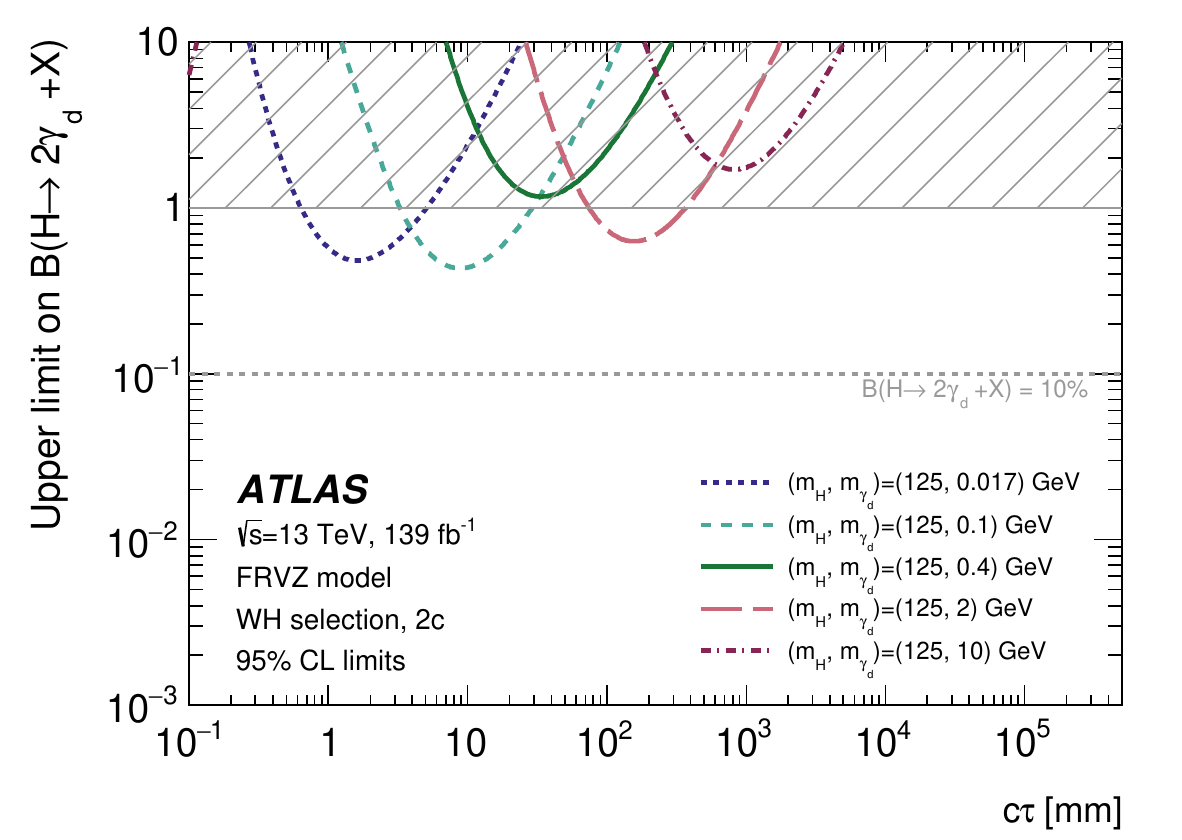}}
\caption{Observed 95$\%$ CL upper limits on the branching ratio ($B$) for the decay \Htwogd and \textit{WH} selection, for different \darky\ masses and a 125~\GeV\ Higgs boson. The limits are shown separately for the three search channels: (a)~c, (b)~c$+\mu$ and (c)~2c, assuming an FRVZ signal model. The hatched band denotes the region in which the branching ratio of \Htwogd is larger than unity.}
\label{fig:limits_allWH}
\end{figure}
 
Figure~\ref{fig:limits_allHAHMggF} reports the observed limits for scenarios assuming the production of a 125~\GeV\ Higgs boson that can decay into two \darky\ in the HAHM model. The sensitivity of each search channel in the ggF selection is reported in a separate subfigure. The sensitivities in the \textit{WH} selection are significantly lower because the CalRatio trigger is less efficient for those events.
For pairs of dark photons decaying into \mLJs, branching fractions larger than 1\% are excluded at 95\% CL if the dark photons have a mean proper decay length \ctau between 4~mm and 200~mm and a mass between 0.4~\GeV\ and 2~\GeV. When considering dark-photon decays into pairs of \hLJs, branching fractions larger than 10\% are excluded at 95\% CL for mean proper decay lengths between 1.5~mm and 8~mm and masses between 17~\MeV\ and 100~\MeV.
 
\begin{figure}[h!]
\centering
\subfloat[\SRggFmm]{\includegraphics[width=.49\textwidth]{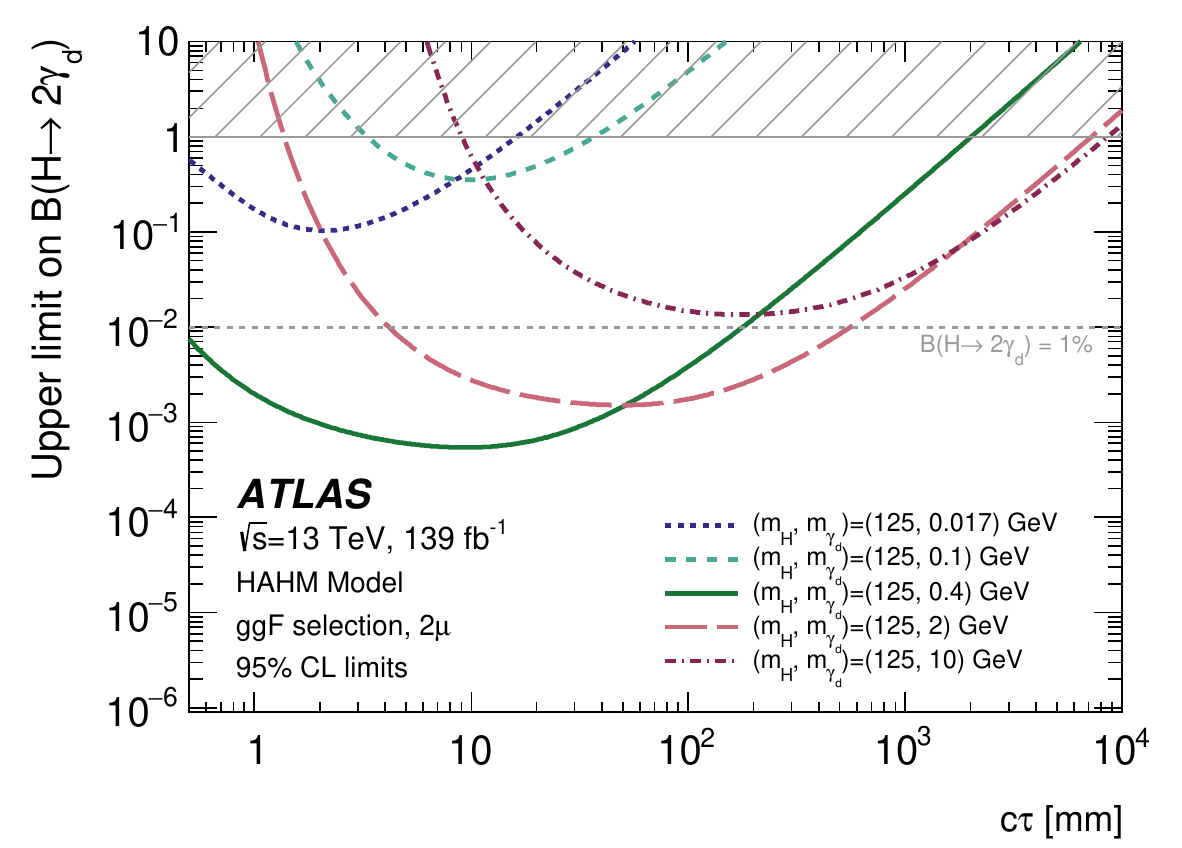}}
\subfloat[\SRggFhm]{\includegraphics[width=.49\textwidth]{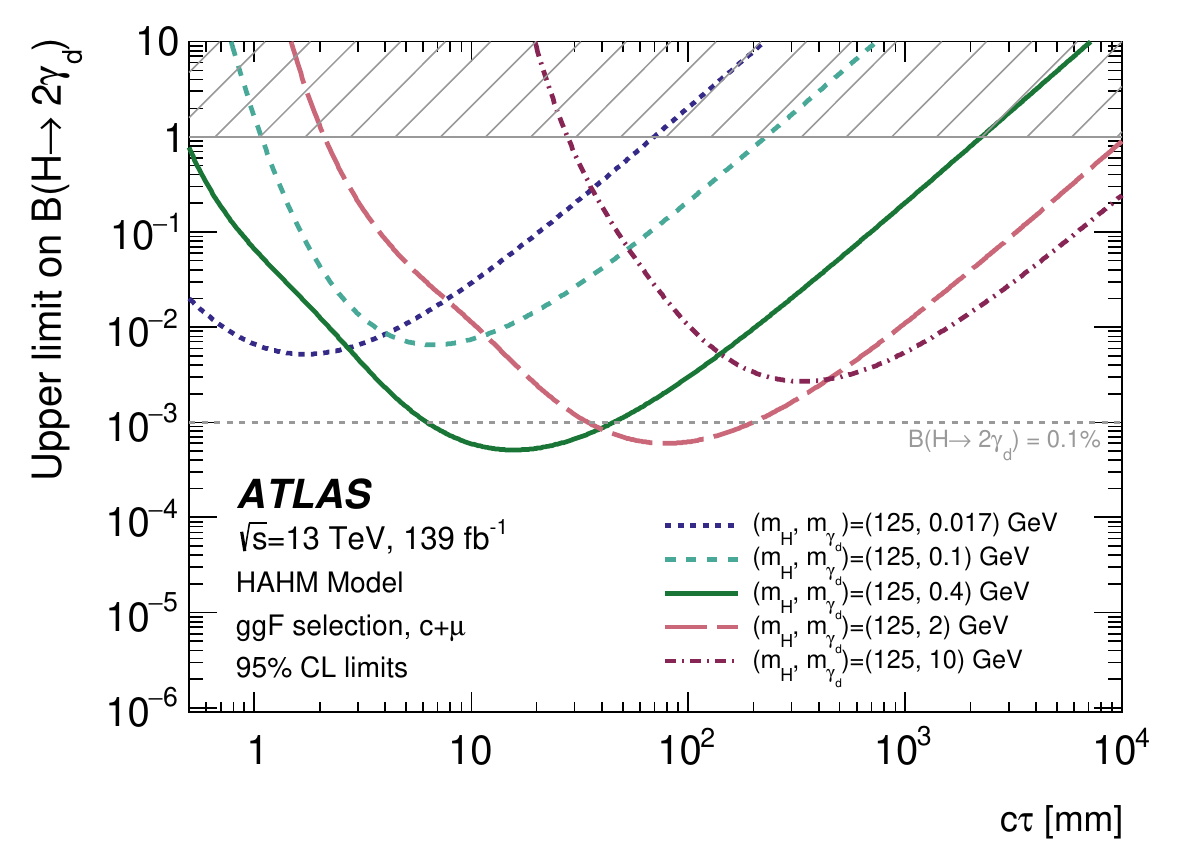}}\\
\subfloat[\SRggFhh]{\includegraphics[width=.49\textwidth]{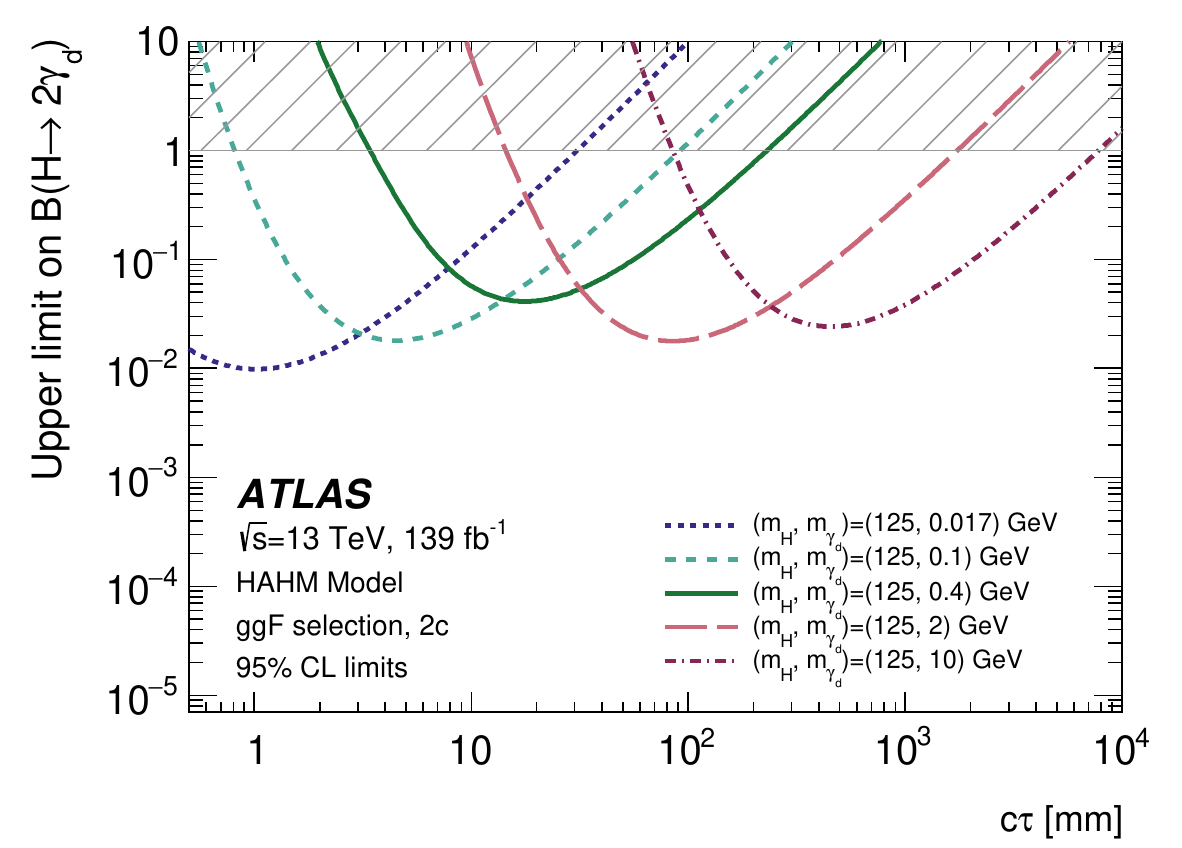}}
\caption{Observed 95$\%$ CL upper limits on the branching ratio ($B$) for the decay \HtwogdHAHM and ggF selection, for different \darky\ masses and a 125~\GeV\ Higgs boson. The limits are shown separately for the (a)~$2\mu$, (b)~c$+\mu$ and (c)~2c search channels, assuming a HAHM signal model. The hatched band denotes the region in which the branching ratio of \HtwogdHAHM is larger than unity.}
\label{fig:limits_allHAHMggF}
\end{figure}
 
Alternative signal scenarios were also considered in the context of FRVZ models.
The sensitivity to scenarios where a pair of dark photons is produced in the decay of an 800~\GeV\ scalar mediator is driven by the selection efficiency being higher than for the decay of the 125~\GeV\ Higgs boson because of the harder dark-photon energy spectrum.
The signal region targeting decays into one \hLJ and one \mLJ in the ggF selection was found to be the most sensitive: branching fractions above 10\% are excluded at 95\% CL if the dark photons have a mean proper decay length between 6~mm and 30~mm and a mass between 400~\MeV\ and 2~\GeV.
In the case of models leading to a total of four dark photons in the final state, the sensitivity is lower than for the \Htwogd process because the large boost of dark-photon pairs from the \sd decays results in \hLJs\ overlapping with \mLJs\ and failing the object-level selections. The signal region targeting decays into one \hLJ and one \mLJ in the ggF selection was found to be the most sensitive: branching fractions above 10\% are excluded at 95\% CL for dark photons with a mean proper decay length between 20~mm and 50~mm and a mass between 400~\MeV\ and 1~\GeV.
In order to set limits on the \gghHtwogd process as a function of the kinetic mixing parameter $\epsilon$ and \darky mass, the best-performing search channels, based on their expected sensitivity, are taken into account. In the case of FRVZ models, the mutually exclusive search channels targeting dark-photon decays into \hLJs are statistically combined to maximise the exclusion power. The combined fit considers a product of the likelihood functions of the individual search channels with independent parameters, but with a common signal normalisation.
Figure~\ref{fig:limits_combi} considers the case of an FRVZ model with a \darky mass of 100~\MeV\ and presents the results of the statistical combination of the \SRggFhh, \SRWHhh and \SRWHh, as 95\% CL upper limits on the branching fraction of a 125~\GeV\ Higgs boson decaying into hidden-sector particles and resulting in two dark photons. For simplicity, the \SRggFhm is not used in the combination although it could offer an increase in sensitivity at the percent level. The figure illustrates the complementary sensitivity of the \textit{WH} selection, in this case providing most of the sensitivity for dark-photon mean proper decay lengths below 2~mm or above 50~mm.

\begin{figure}[h!]
\centering
\includegraphics[width=.49\textwidth]{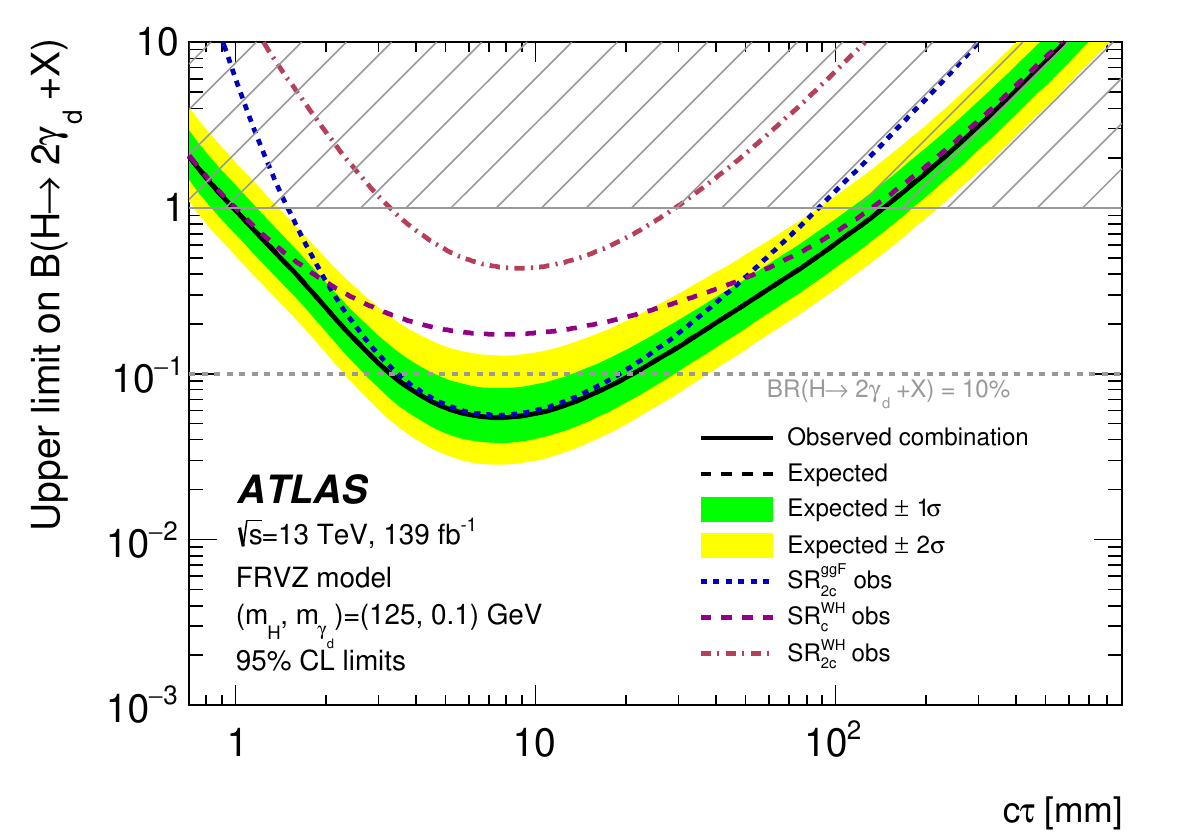}
\caption{Exclusion limits at the 95$\%$ CL on the branching ratio ($B$) for the process \Htwogd, for a \darky\ mass of 100~\MeV\ and a 125~\GeV\ Higgs boson in an FRVZ model, as a function of the dark-photon mean proper decay length \ctau. The three dashed curves, taken from Figures~\ref{fig:limits_allggF}(c), \ref{fig:limits_allWH}(a) and \ref{fig:limits_allWH}(c), show the individual sensitivities of the search channels targeting \hLJs in the ggF and \textit{WH} selections. The solid black curve shows the observed exclusion limit from their statistical combination, and is almost identical to the expected limit. The green and yellow bands represent $\pm 1\sigma$ and $\pm 2\sigma$ uncertainty in the expected limit. The hatched band denotes the region in which the branching ratio of \Htwogd is larger than unity.
}
\label{fig:limits_combi}
\end{figure}
 
Upper limits at 90\% CL\footnote{The limits are quoted at 90\% CL in this interpretation of the results to ease the comparison with other experiments that conventionally use this CL threshold.} are also set, in the context of the FRVZ model and the HAHM vector-portal model, in terms of kinetic mixing parameter $\epsilon$ and \darky mass and presented in Figure~\ref{fig:DMexclusionNew} for different Higgs decay branching fractions into \darky, ranging from 0.1\%~to~10\%.
The limits are interpolated between different masses by branching fraction variations~\cite{Batell:2009yf} as a function of the \darky mass, corrected by a linear interpolation of the signal efficiency between adjacent available MC signal samples.
For \darky mass below twice the muon mass, no coverage from \mLJ signal regions is expected, motivating a significant drop in sensitivity. The combination of the \hLJ signal regions provides the most stringent limit in this mass range.
In the mass range above twice the muon mass, only the best-performing channel for each \darky mass probed is considered.
The structures observed in this mass range depend on the expected \darky branching ratio, where decays to QCD resonances can significantly alter the sensitivities of the different search channels.
 
Exclusion regions from previous ATLAS searches for dark-photon jets are shown for the displaced-decay search~\cite{EXOT-2017-28} and for the complementary prompt-decay search~\cite{EXOT-2014-09}. The contributions from the \hLJ channels allow ATLAS to set limits for $m_{\gamma_{d}}<0.1\,\gev$ for the first time.
 
\begin{figure}[ht]
\centering
\subfloat[FRVZ]{\includegraphics[width=.45\textwidth]{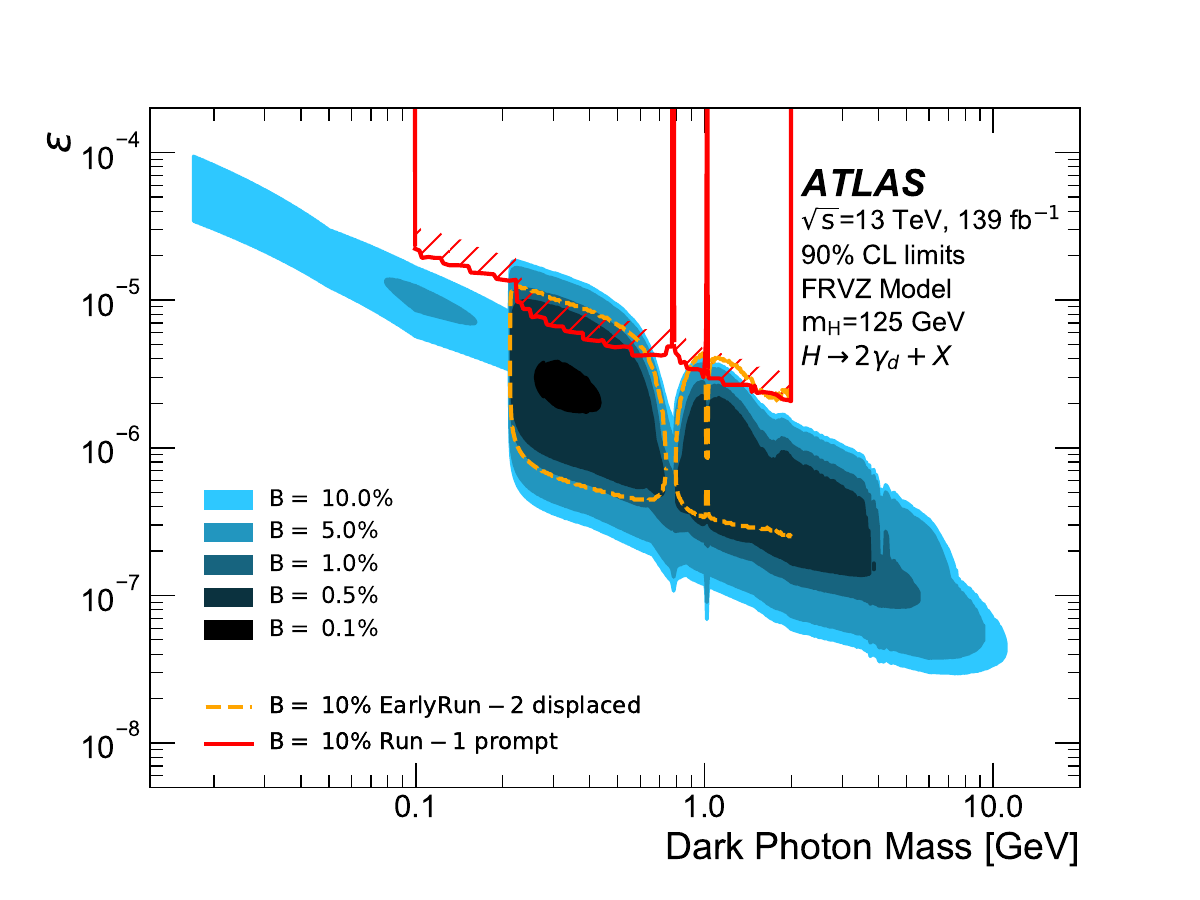}}
\subfloat[HAHM]{\includegraphics[width=.45\textwidth]{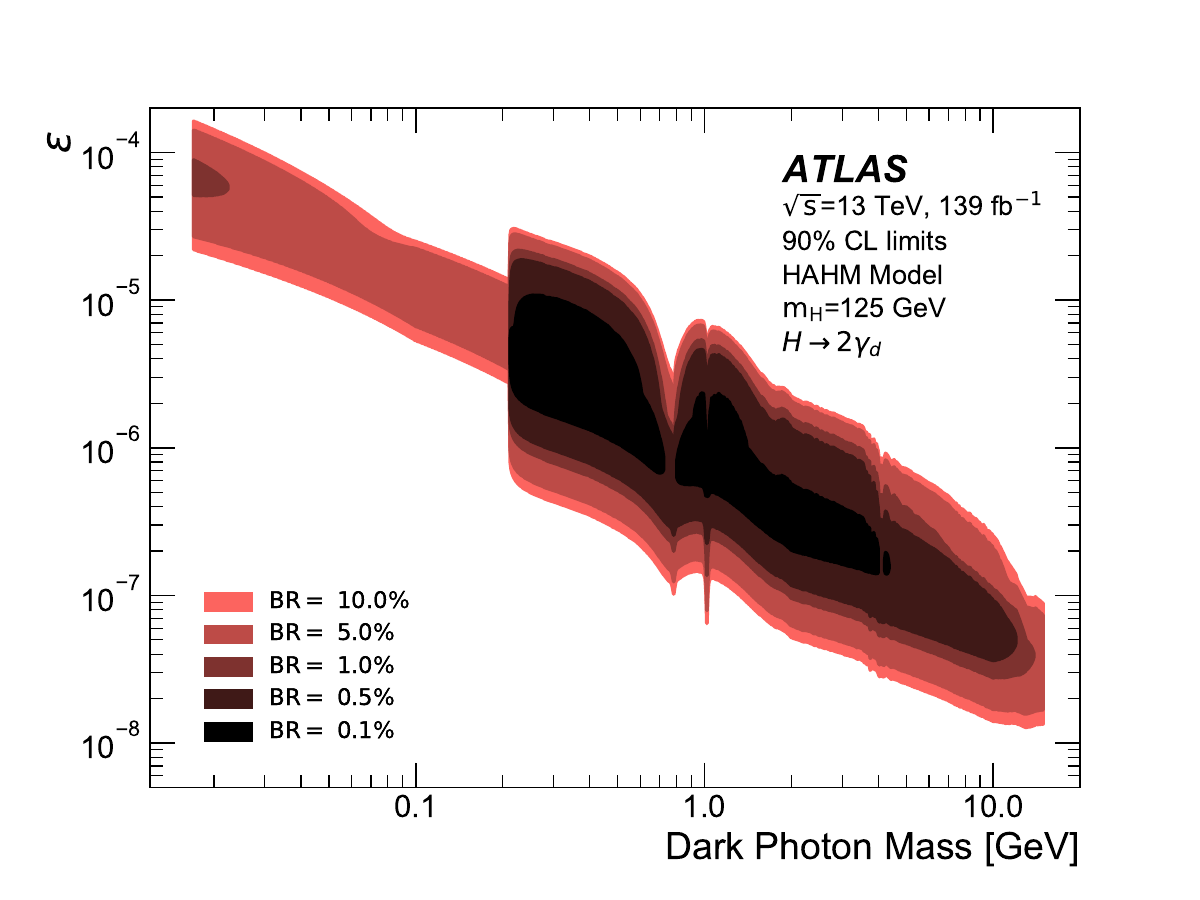}}
\caption{The 90$\%$ CL exclusion contours of the branching ratio ($B$) for the decay \Htwogd as a function of the \darky\ mass and the kinetic mixing parameter $\epsilon$ for (a) the FRVZ model and (b) the HAHM model.
These limits are obtained assuming branching fractions between 0.1\% and 10\% for Higgs boson decays resulting in dark photons.
For \darky~mass below twice the muon mass, the combination of the \hLJ signal regions is used.
The figure also shows regions excluded by the previous ATLAS searches for jets from displaced~\cite{EXOT-2017-28} (orange line) and prompt~\cite{EXOT-2014-09} (red line) decays of dark photons.
}
\label{fig:DMexclusionNew}
\end{figure}
 
This search has significantly higher sensitivity than the previous ATLAS search for light long-lived neutral particles~\cite{EXOT-2017-28}, reaching a 0.1\% branching ratio for a 125~\GeV\ Higgs boson decaying into dark photons.


\FloatBarrier

\section{Conclusion}
\label{sec:conclusion}
 
This paper presents a search for light long-lived neutral particles which decay into collimated pairs of fermions in the ATLAS detector at the LHC. Data collected from 139~fb$^{-1}$ of $\sqrt{s}=13$~\TeV\ $pp$ collisions during 2015--2018 were analysed for evidence of long-lived dark photons from Higgs boson decays and were found to be consistent with the background prediction. Upper limits on the Higgs boson branching fraction to dark photons as a function of their mass and mean proper decay length \ctau are reported, assuming SM Higgs boson production cross-sections. Branching fractions above 1\% are excluded at 95\% CL for Higgs boson decays to two dark photons with mean proper decay length between 10~mm and 250~mm and mass between 0.4 and 2~\GeV. This analysis has significantly higher sensitivity than previous ATLAS searches for light long-lived neutral particles that decay to collimated pairs of fermions. The higher sensitivity is due to the addition of an event selection targeting long-lived dark photons from decays of Higgs bosons produced in association with a $W$ boson and to the use of a larger dataset and new methodologies for reconstructing dark-photon candidates and rejecting non-collision backgrounds.


\FloatBarrier
 
\section*{Acknowledgements}


We thank CERN for the very successful operation of the LHC, as well as the
support staff from our institutions without whom ATLAS could not be
operated efficiently.
 
We acknowledge the support of
ANPCyT, Argentina;
YerPhI, Armenia;
ARC, Australia;
BMWFW and FWF, Austria;
ANAS, Azerbaijan;
CNPq and FAPESP, Brazil;
NSERC, NRC and CFI, Canada;
CERN;
ANID, Chile;
CAS, MOST and NSFC, China;
Minciencias, Colombia;
MEYS CR, Czech Republic;
DNRF and DNSRC, Denmark;
IN2P3-CNRS and CEA-DRF/IRFU, France;
SRNSFG, Georgia;
BMBF, HGF and MPG, Germany;
GSRI, Greece;
RGC and Hong Kong SAR, China;
ISF and Benoziyo Center, Israel;
INFN, Italy;
MEXT and JSPS, Japan;
CNRST, Morocco;
NWO, Netherlands;
RCN, Norway;
MEiN, Poland;
FCT, Portugal;
MNE/IFA, Romania;
MESTD, Serbia;
MSSR, Slovakia;
ARRS and MIZ\v{S}, Slovenia;
DSI/NRF, South Africa;
MICINN, Spain;
SRC and Wallenberg Foundation, Sweden;
SERI, SNSF and Cantons of Bern and Geneva, Switzerland;
MOST, Taiwan;
TENMAK, T\"urkiye;
STFC, United Kingdom;
DOE and NSF, United States of America.
In addition, individual groups and members have received support from
BCKDF, CANARIE, Compute Canada and CRC, Canada;
PRIMUS 21/SCI/017 and UNCE SCI/013, Czech Republic;
COST, ERC, ERDF, Horizon 2020 and Marie Sk{\l}odowska-Curie Actions, European Union;
Investissements d'Avenir Labex, Investissements d'Avenir Idex and ANR, France;
DFG and AvH Foundation, Germany;
Herakleitos, Thales and Aristeia programmes co-financed by EU-ESF and the Greek NSRF, Greece;
BSF-NSF and MINERVA, Israel;
Norwegian Financial Mechanism 2014-2021, Norway;
NCN and NAWA, Poland;
La Caixa Banking Foundation, CERCA Programme Generalitat de Catalunya and PROMETEO and GenT Programmes Generalitat Valenciana, Spain;
G\"{o}ran Gustafssons Stiftelse, Sweden;
The Royal Society and Leverhulme Trust, United Kingdom.
 
The crucial computing support from all WLCG partners is acknowledged gratefully, in particular from CERN, the ATLAS Tier-1 facilities at TRIUMF (Canada), NDGF (Denmark, Norway, Sweden), CC-IN2P3 (France), KIT/GridKA (Germany), INFN-CNAF (Italy), NL-T1 (Netherlands), PIC (Spain), ASGC (Taiwan), RAL (UK) and BNL (USA), the Tier-2 facilities worldwide and large non-WLCG resource providers. Major contributors of computing resources are listed in Ref.~\cite{ATL-SOFT-PUB-2021-003}.


\clearpage
\printbibliography
 
\clearpage
 
\begin{flushleft}
\hypersetup{urlcolor=black}
{\Large The ATLAS Collaboration}

\bigskip

\AtlasOrcid[0000-0002-6665-4934]{G.~Aad}$^\textrm{\scriptsize 101}$,
\AtlasOrcid[0000-0002-5888-2734]{B.~Abbott}$^\textrm{\scriptsize 119}$,
\AtlasOrcid[0000-0002-7248-3203]{D.C.~Abbott}$^\textrm{\scriptsize 102}$,
\AtlasOrcid[0000-0002-2788-3822]{A.~Abed~Abud}$^\textrm{\scriptsize 36}$,
\AtlasOrcid[0000-0002-1002-1652]{K.~Abeling}$^\textrm{\scriptsize 55}$,
\AtlasOrcid[0000-0002-2987-4006]{D.K.~Abhayasinghe}$^\textrm{\scriptsize 94}$,
\AtlasOrcid[0000-0002-8496-9294]{S.H.~Abidi}$^\textrm{\scriptsize 29}$,
\AtlasOrcid[0000-0002-9987-2292]{A.~Aboulhorma}$^\textrm{\scriptsize 35e}$,
\AtlasOrcid[0000-0001-5329-6640]{H.~Abramowicz}$^\textrm{\scriptsize 150}$,
\AtlasOrcid[0000-0002-1599-2896]{H.~Abreu}$^\textrm{\scriptsize 149}$
% ...
\bigskip
\\

% ...
$^{29}$Physics Department, Brookhaven National Laboratory, Upton NY; United States of America.\\
% ...
$^{35}$$^{(a)}$Facult\'e des Sciences Ain Chock, R\'eseau Universitaire de Physique des Hautes Energies - Universit\'e Hassan II, Casablanca;$^{(b)}$Facult\'{e} des Sciences, Universit\'{e} Ibn-Tofail, K\'{e}nitra;$^{(c)}$Facult\'e des Sciences Semlalia, Universit\'e Cadi Ayyad, LPHEA-Marrakech;$^{(d)}$LPMR, Facult\'e des Sciences, Universit\'e Mohamed Premier, Oujda;$^{(e)}$Facult\'e des sciences, Universit\'e Mohammed V, Rabat;$^{(f)}$Institute of Applied Physics, Mohammed VI Polytechnic University, Ben Guerir; Morocco.\\
$^{36}$CERN, Geneva; Switzerland.\\
% ...
$^{55}$II. Physikalisches Institut, Georg-August-Universit\"{a}t G\"ottingen, G\"ottingen; Germany.\\
% ...
$^{94}$Department of Physics, Royal Holloway University of London, Egham; United Kingdom.\\
% ...
$^{101}$CPPM, Aix-Marseille Universit\'e, CNRS/IN2P3, Marseille; France.\\
$^{102}$Department of Physics, University of Massachusetts, Amherst MA; United States of America.\\
% ...
$^{119}$Homer L. Dodge Department of Physics and Astronomy, University of Oklahoma, Norman OK; United States of America.\\
% ...
$^{149}$Department of Physics, Technion, Israel Institute of Technology, Haifa; Israel.\\
$^{150}$Raymond and Beverly Sackler School of Physics and Astronomy, Tel Aviv University, Tel Aviv; Israel.\\
% ...
$^{a}$ Also Affiliated with an institute covered by a cooperation agreement with CERN.\\
$^{b}$ Also at Borough of Manhattan Community College, City University of New York, New York NY; United States of America.\\
$^{c}$ Also at Bruno Kessler Foundation, Trento; Italy.\\
$^{d}$ Also at Center for High Energy Physics, Peking University; China.\\
$^{e}$ Also at Centro Studi e Ricerche Enrico Fermi; Italy.\\
$^{f}$ Also at CERN, Geneva; Switzerland.\\
% ...

\end{flushleft}


\end{document}